\documentclass[letterpaper,oneside,10pt]{article}
\usepackage[USenglish]{babel} 
\usepackage[ansinew]{inputenc}
\usepackage{amsmath, amsthm, amssymb, amsfonts}
\usepackage{verbatim}
\usepackage{graphicx}
\usepackage{color}
\usepackage{float}
\usepackage{latexsym}
\usepackage{amsmath}
\usepackage{amsfonts}
\usepackage{enumerate}
\usepackage{listings}
\usepackage{caption}
\usepackage{subcaption}
\usepackage{url}
\usepackage[margin=0.62in,footskip=0.2in]{geometry}

\renewcommand{\Re}{\mathbb{R}}

\newcommand{\benum}{\begin{enumerate}}
\newcommand{\eenum}{\end{enumerate}}
\newcommand{\nc}{\newcommand}
\newcommand{\rnc}{\renewcommand}
\newcommand{\bvb}{\begin{verbatim}}
\newcommand{\evb}{\end{verbatim}}

\DeclareMathOperator{\Co}{Co}

\nc{\bq}{\textbf}
\nc{\m}{\textrm}
\nc{\bb}{\mathbb}
\rnc{\bf}{\mathbf}
\nc{\til}{\texttildelow}
\nc{\be}{\begin{equation}}
\nc{\ee}{\end{equation}}
\nc{\dps}{\displaystyle}
\rnc{\l}{\left(}\rnc{\r}{\right)}
\nc{\lc}{\left\{}\nc{\rc}{\right\}}
\nc{\lb}{\left[}\nc{\rb}{\right]}
\nc{\ba}[1]{\begin{array}{#1}}
\nc{\ea}{\end{array}}
\nc{\ra}{\rightarrow}
\nc{\li}{\left |}
\nc{\ri}{\right |}
\nc{\pde}[2]{\frac{\partial #1}{\partial #2}}
\nc{\ode}[2]{\frac{d #1}{d #2}}
\nc{\odee}[3]{\frac{d^{#3} #1}{d #2^{#3}}}
\nc{\pdee}[3]{\frac{\partial^{#3} #1}{\partial #2^{#3}}}
\nc{\bn}{\begin{enumerate}}
\nc{\en}{\end{enumerate}}
\nc{\bt}{\begin{theorem}}
\nc{\et}{\end{theorem}}
\nc{\y}[1]{\lambda_{#1}}
\nc{\ir}{\mathbb{I}\mathbb{R}}

\nc{\ep}{\mathcal{E}_{P}}
\nc{\mr}{\mathcal{M}_{r}}
\nc{\mfa}{\mathcal{M}_{f,a}}
\nc{\mfp}{\mathcal{M}_{f,p}}
\nc{\mt}{\m{T}}
\nc{\F}{\mathbb{F}}

\begin{document}

\newtheorem{proposition}{Proposition}
\newtheorem{thm}{Theorem}
\newtheorem{rmk}{Remark}

\definecolor{cgreen}{rgb}{0,0.6,0}
\lstset{ %
language=C,                
basicstyle=\footnotesize\ttfamily,       
numbers=left,                   
numberstyle=\scriptsize,      
stepnumber=1,                   
numbersep=4pt,                  
commentstyle=\color{cgreen},
backgroundcolor=\color{white},  
showspaces=false,               
showstringspaces=false,         
showtabs=false,                 
frame=single,   		
tabsize=2,  		
captionpos=b,   		
breaklines=true,    	
breakatwhitespace=false,    
escapeinside={\%}{)}          
}

\pagestyle{plain}

\title{\textbf{Verifiable Control System Development for Gas Turbine Engines}}

\author{Mehrdad Pakmehr \footnote{Postdoctoral fellow at the School of Aerospace Engineering, Georgia Institute of Technology, Atlanta, GA 30332, {\small mehrdad.pakmehr@gatech.edu.}},
Timothy Wang \footnote{PhD Candidate at the School of Aerospace Engineering, Georgia Institute of Technology, Atlanta, Georgia 30332, {\small timothy.wang@gatech.edu.}},
Romain Jobredeaux \footnote{PhD Candidate at the School of Aerospace Engineering, Georgia Institute of Technology, Atlanta, Georgia 30332, {\small romain.jobredeaux@gatech.edu.}},
Martin Vivies \footnote{System Engineer at the Price Induction Inc., Marietta, Georgia, 30067, {\small martin.vivies@price-induction.com.}},
Eric Feron \footnote{Professor at the School of Aerospace Engineering, Georgia Institute of Technology, Atlanta, GA 30332, {\small feron@gatech.edu.}}      }

\date{\null}

\maketitle
\pagestyle{plain} 

\section*{Abstract}

A control software verification framework for gas turbine engines is developed. A stability proof is presented for gain scheduled closed-loop engine system based on global linearization and linear matrix inequality (LMI) techniques. Using convex optimization tools, a single quadratic Lyapunov function is computed for multiple linearizations near equilibrium points of the closed-loop system. With the computed stability matrices, ellipsoid invariant sets are constructed, which are used efficiently for DGEN turbofan engine control code stability analysis. Then a verifiable linear gain scheduled controller for DGEN engine is developed based on formal methods, and tested on the engine virtual test bench. Simulation results show that the developed verifiable gain scheduled controller is capable of regulating the engine in a stable fashion with proper tracking performance.

\section*{Nomenclature}

$NL$: Low Pressure Spool Speed\\
$NH$: High Pressure Spool Speed\\
$W_f$: Fuel Flow Control Input\\
PLA: Power Lever Angle\\
SFC: Specific Fuel Consumption\\
FADEC: Full Authority Digital Engine Control\\
WESTT: Whole Engine Simulator Turbine Technology\\
ECU: Engine Control Unit\\
$\alpha$: Scheduling Parameter\\
\emph{Co}: Convex Hull\\
superscript \emph{c}: Controller\\
superscript \emph{p}: Plant\\
superscript \emph{ol}: Open-loop\\
superscript \emph{cl}: Closed-loop\\

\section{Introduction}

Stability and control of gas turbine engines have been of interest to researchers and engineers from a variety of perspectives. Some of the literature related to engine control can be found in \cite{powerplantControl-sobey-1963, control-spang-1999, enginecontrol-jaw-2009, robustAeroengine-arriffin-1997, lqg-athans-1986, lqg-garg-1989, turbofanControl-fredrick-2000, AdvControl-richter-2012, PhDThesis-pakmehr-2013}. To facilitate the stability analysis of nonlinear systems, such as gas turbine engines, an efficient technique is to approximate them by a linear time-varying (LTV) system. One of the control design approaches, which perhaps is one of the most popular nonlinear control design approaches and has been widely and successfully applied in fields ranging from aerospace to process control \cite{research-rugh-2000, surveyGS-leith-2000}, is gain scheduling. Gas turbine engines are no exception, and research on gain scheduling of gas turbine engines is presented in \cite{kapasouris-gainsched-1985, turbofanSched-garg-1997, lpv-bruzelius-2002, lpv-balas-2002, lpv-gilbert-2010, approximate-zhao-2011, GainSchedStabConf-pakmehr-2013, GSstability-pakmehr-2013}.

When the operation of a control system is highly critical due to human safety factors or the high cost of failure in damaged capital or products, the software designers have to expend more effort to validate and verify their software before it can be released. In flight-critical operations, validation and verification are part of the flight certification process \cite{SoftwareTech-heck-2003}. Software system certification involves many challenges, including the necessity to certify the system at the level of functional requirements, code and binary levels, the need to detect run-time errors, and the need for proving timing properties of the eventual, compiled system \cite{CertifyCont-feron-2007, DO178C-RTCA-2011}.

Provable closed-loop stability constitutes an essential attribute of control systems, especially when human safety is involved, as in many aeronautical systems like gas turbine engines. Motivated by such applications, there exist many theorems to support system stability and performance under various assumptions \cite{ContSoftware-feron-2010}. Stability criteria apply to a class of dynamical systems for which a stability proof is established; and Lyapunov's stability theory plays a critical role in this regard. Control-system domain knowledge, in particular, Lyapunov-theoretic proofs of stability and performance, can be migrated toward computer-readable and verifiable certificates \cite{ContSoftware-feron-2010, autocoding-feron-2011}. Some of the recent research results on the control software verification using Lyapunov proof of stability can be reviewed in \cite{CertifyCont-feron-2007, ContSoftware-feron-2010, autocoding-feron-2011, ProofCode-Jobredeaux-2012, PhdThesis-roozbehani-2008, LyapSoftVerif-roozbehani-2013, GraphEnviron-wang-2011, DesToImpAutomated-wang-2013}.

Software verification process for aerospace systems is explained in ``RTCA /DO-178B: Software Considerations in Airborne Systems and Equipment Certification" \cite{DO178B-RTCA-1992}. Currently there is no detailed theoretical process for software verification; and the verification is mainly performed by running the software long enough to make sure that it works properly for the system at hand. Since the publication of DO-178B \cite{DO178B-RTCA-1992}, experience and scientific advances have been gained in the formal methods, their application, and tools. Formal methods are mathematically based techniques for specification, development, and verification of software aspects of digital systems \cite{DO333-RTCA-2011}. ``RTCA/DO-333: Formal Methods Supplement to DO-178C and DO-278A" \cite{DO333-RTCA-2011} provides guidance for applicants to facilitate the use of formal methods in aerospace systems.

In this paper, we aim at taking the first steps towards a more rigorous software verification process for gas turbine engine control systems, by developing stability proofs for the entire engine control architecture using the Lyapunov stability theory. This approach helps us in constructing an ellipsoid invariant set \cite{lmi-boyd-1994, ellipsoidalCalc-kurzhanski-1997} to be used as an efficient tool for control code stability analysis.

The rest of this paper is organized as follows. Section 2 describes the DGEN 380 turbofan engine and its virtual test bench. Section 3 presents the credible autocoding and verification process for the controller code. Section 4 presents the verifiable controller design and development for the DGEN engine. Section 5 presents the stability analysis for the closed-loop engine system with its gain scheduled controller. Section 6 presents some of the output verifiable code from the credible autocoding process. Section 7 presents simulation results of the verifiable controller code executed on the engine virtual test bench. Section 8 concludes the paper.

\section{Turbofan Engine}

The gas turbine engine model used in this study is a high fidelity model of DGEN 380 turbofan engine provided as a virtual test bench by Price Induction \cite{dgen380-link}. Brief descriptions of the DGEN engine and its virtual test bench are given below.

\subsection{Price Induction DGEN 380 Turbofan Engine}
Price Induction, a French aerospace company based out of Biarritz, has been designing for the past 10 years a family of engines designed specifically for general aviation. These DGEN engines are optimized for a cruise altitude ranging from 15,000 to 20,000 ft, at speed up to Mach 0.35 and with a flight ceiling limited to 25,000 ft. To be competitive with piston engines, they were also designed to have low specific fuel consumption (SFC), be lightweight and reasonably priced to allow the emergence of 4 to 5-seat light aircraft with a max weight ranging between 1,400 kg and 2,150 kg (3,417 lb and 5,622 lb) commonly called Personal Light Jets (PLJ).

\begin{figure}[!ht]
\centering
\includegraphics[width=0.5\textwidth]{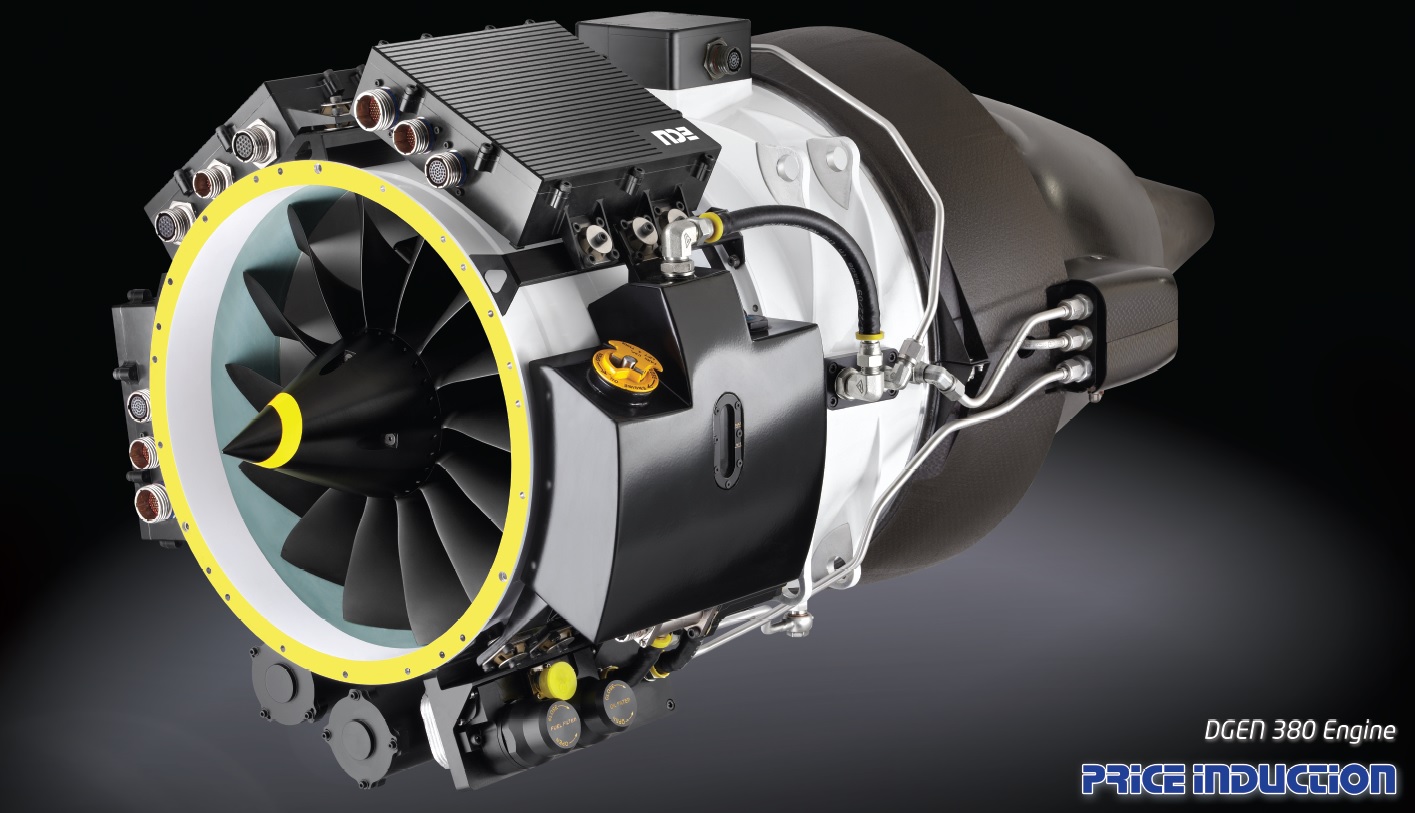}
\caption{DGEN 380 lightweight turbofan engine (\copyright Price Induction)\cite{dgen380-link}}\label{fig:dgen}
\end{figure}

The DGEN 380, the first engine of its family shown in Figure \ref{fig:dgen}, is a two-spool, high bypass ratio (7.6), unmixed flow turbofan engine. Its simple architecture yields up to 560 pounds of thrust in a compact and lightweight format (175 pounds and 4 feet long) while maintaining low noise and pollution levels. Beside its optimized performances, the engine innovates with its all-electric system: its starter-generator located directly on the high-pressure shaft, and oil and fuel pumps driven by electric motors are controlled by the Engine Control Unit (ECU), allowing for a really fine and optimized tuning of the DGEN control laws.

\subsection{Price Induction Engine Virtual Test Bench}
Based on its design expertise, Price Induction developed the WESTT (Whole Engine Simulator Turbine Technology) Solutions, which are educational and research tools based on the DGEN 380 turbofan engine and intended for high schools, universities, aeronautical maintenance training centers and research institutes.

\begin{figure}[!ht]
\centering
\includegraphics[width=0.7\textwidth]{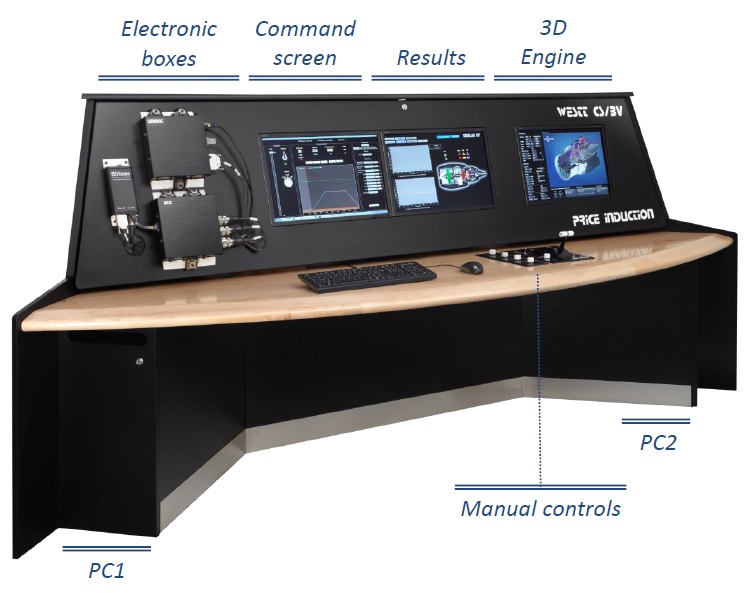}
\caption{ Price Induction WESTT CS-BV: DGEN 380 turbofan engine virtual test bench (\copyright Price Induction) \cite{dgen380-link}} \label{fig:westt}
\end{figure}

The WESTT CS-BV, shown in Figure \ref{fig:westt}, is a product of this family dedicated to the study of the DGEN 380 turbofan and its control. With the DGEN 380 actual ECU hardware and its model running real-time and generating its sensors analog outputs, the CS-BV constitutes a great control Hardware-In-the-Loop (HIL) platform for the testing of engine control design. The MPC555 microcontroller which constitutes the core of the ECU can be easily programmed through the already existing code framework with different control logics and tested in real time with the use of the SIMMOT (engine real-time simulation). All engine outputs are displayed on screen and all data recorded for later performance analysis.

\section{Credible Autocoding and Verification Process}\label{autocodeProcess}

\begin{figure}[!ht]
\centering
\includegraphics[width=0.7\textwidth]{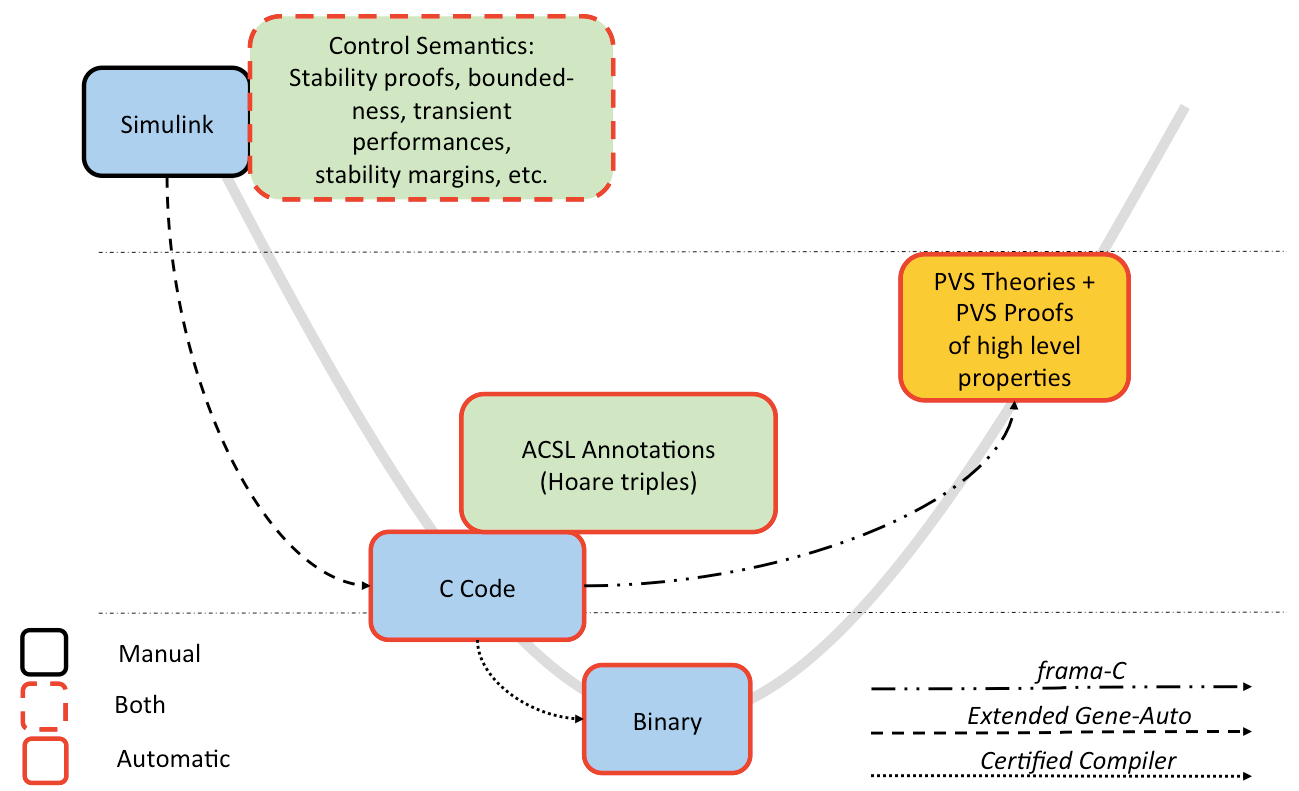}
\caption{Visualization of autocoding and verification process} \label{fig:verified_process}
\end{figure}

The Lyapunov matrices which are computed in the next sections can be used at the level of the code in order to formally verify its stability. Indeed the sublevel sets of the Lyapunov functions they represent are invariant sets in which the variables remain throughout the execution of the program. The concept of invariants is familiar to the computer scientists that seek them in computer programs. However, a general computer program lacks the nice mathematical structures that exist in control systems, which makes it hard to compute invariants for the general computer code.  Instead, we target control systems by developing a control domain-specific framework of \emph{credible autocoding} and implementing it using a number of existing tools that we have extended. A high level view of the credible autocoding framework is given in Figure \ref{fig:verified_process}. It enables the control engineer to provide the control semantics such as the proof of Lyapunov stability directly on the Simulink diagram. This part of the framework of can be both manual or automated. The implementation of the framework is done by extending an Simulink to C translation tool named Gene-Auto \cite{nassimaformal09} to generate formal annotations in additional to the C code. These annotation describe the invariant sets in which the variables will remain. The sets are extracted automatically from the Lyapunov matrices. The formal annotation language used at the level of the C code is called ACSL (ANSI/ISO C Specification Language) \cite{acsl-link}. It can be read by various formal analysis tools such as Frama-C \cite{framaC-link}. In order to prove the validity of the generated annotations, we have extended these tools with domain-specific, automated routines that discharge the proof based on control-theoretic techniques. 

\section{Verifiable Engine Controller Design}

The FADEC used for numerical analysis is a DGEN 380 turbofan engine controller model provided by Price Induction. The main novelty with this FADEC is that it is intended to be adaptable to different components such as a different compressor or turbine. The engine controller is developed using gain scheduling, and the scheduling parameter is the engine high pressure spool speed, i.e. $\alpha=NH$.

\subsection{Initial Model and Alterations to the Model}
The initial model requires extensive changes before it become compatible with our tool-chain. The majority of the changes required are not due to the Simulink blocks but rather because of the heavy presence of Matlab functions in the model. A large portion of modifications are made to the parts of the model that is not directly related to the controller.  The model is consisted of three major subsystems. The first major subsystem is the throttle subsystem. This subsystem generates turbine commands NH and NL from the throttle input PLA. This subsystem is mostly written in Matlab and therefore need to be completely re-implemented in Simulink. The second major subsystem is the controller which contained transfer functions, and a polynomial curves-based lookup table. Both type of blocks are not supported by our tool-chain hence need to be converted into subsystems consisted of more basic blocks. The overall controller is consisted of dual PID subsystems with a basic anti-windup mechanism on both PID controllers. The last major subsystem is the Butee which contained safety features such as saturation operators and min/max switching strategies that are designed to prevent controller commands that could cause the turbine to exceed its maximum operational safety limits. This part of the model also requires extensive alteration as most of the Butee is written in Matlab.

\subsection{Analysis}
In this section, we give a description of the open-loop stability characteristics of the FADEC. These results are to be used to annotate the Simulink model and then transformed by our credible compilation tool-chain down to ACSL annotations for the C code.

From the Simulink model of FADEC provided by Price Induction, the semantics of the controller is extracted and then reformulated as a discrete-time state-space system. Let the controller states be denoted by the vector $x^c \in \Re^{11}$. The controller states are
\be
\dps x^c= \lb \ba{ccccccccccc} b_{0} & b_{1} & \epsilon_{0} & \epsilon_{1} & c_{0} & c_{1} & f_{0} & f_{1} & b_{2} & \epsilon_{2} & c_{2} \ea \rb^{\mathsf{T}}.
\label{controller_states}
\ee
The symbols $b_{n},n=0,1,2$ represent the integrator states, and the symbols $\epsilon_{n}, c_{n}, n=0,1,2 $ denote the states of the anti-windup mechanisms. The states $f_{n},n=0,1$ are the states of the first order filters used in the derivative portion of the two PID controllers.

We have $y_{1} \in \Re^{2}$ as one of the input to the controller
\be
\dps y_{1} =\lb \ba{c} \Delta NH  \cr \Delta NL \ea\rb
\label{input:01}
\ee.
The symbols $\Delta NH$ and $\Delta NL$ are respectively the changes to the high and low pressure turbine spool speeds commanded by the throttle subsystem. The output from the controller, denoted as $u \in \Re$, is the input to the Butee subsystem. The Butee component is  a safety limiter on the signal $u$. The output from the Butee, denoted as $\hat{u}$, is the input to the engine fuel pump. There is a feedback loop to the controller. This feedback loop to the controller contains two signals. One is the output from the Butee $\hat{u}$ and the other is $u_{2}$ which is a vector consisted of the anti-windup states $c_{n},n=1,2,3$. Both signals are delayed by one sample period in the feedback loop.

\begin{figure}[!ht]
\centering
\includegraphics[scale=0.28]{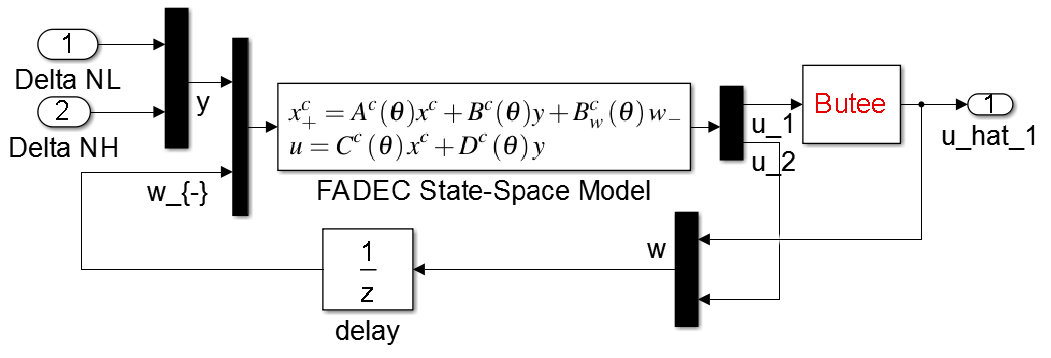}
\caption{State-space model of the FADEC developed in Simulink}
\label{fig:fadec}
\end{figure}

The symbols $NH$ and $NL$ denote the angular velocity of the high-pressure and low-pressure spools.  The PID controller gains are computed using polynomial functions $p_{n},n=1,\ldots,4$ that map $NH$ to a set of gains
\be \ba{c}
\dps p_{1}: NH \ra K_{p} \cr
\dps p_{2}: NH \ra K_{i} \cr
\dps p_{3}: NH \ra K_{d} \cr
\dps p_{4}: NH \ra T_{d}.
\ea
\label{gains}
\ee
Let the parameter
\be
\dps \theta(\alpha) =\lb \ba{cccc} p_{1}(\alpha) & p_{2}(\alpha) & p_{3}(\alpha) & p_{4}(\alpha)  \ea \rb
\label{parameter}
\ee
denote the set of PID gains for some $\alpha$. The state-space transition function and the output function of the FADEC can be defined using the following parameter-varying matrices.
\be \ba{llll}
\dps A^c(\theta) \in \Re^{11 \times 11}, & B^c(\theta) \in \Re^{11 \times 2}, & B^c_{u_{2}}(\theta) \in \Re^{11 \times 3},  & B^c_{\hat{u}}(\theta) \in \Re^{11 \times 1} \cr
\dps C^c_{1}(\theta) \in \Re^{1 \times 11}, & C^c_{2}(\theta) \in \Re^{3 \times 11}, & D^c_{1}(\theta) \in \Re^{1 \times 2 }
\ea
\label{abcd}
\ee
The matrices are varying in a nonlinear fashion with the parameter $\theta$. Let $B^c_{w}(\theta)  =\lb \ba{ccc} B^c_{\hat{u}_{1}}(\theta) & B^c_{u_{2}}(\theta) \ea \rb$, $C^c(\theta)=\lb \ba{c} C^c_{1}(\theta) \cr C^c_{2}(\theta) \ea \rb$, $D^c(\theta)=\lb \ba{c} D^c_{1}(\theta) \cr 0 \ea \rb$, $u_{1}=C^c_{1}(\theta) x^c + D^c_{1}(\theta) y$, $u_{2}=C^c_{2}(\theta) x^c$, $\hat{u}_{1} =\sigma(u_{1})$, where $\sigma$ is the nonlinear causual operator representing the Butee, and finally let $w = \lb \ba{c} \hat{u}_{1}  \cr u_{2} \ea \rb$. The discrete-time linear state-space model of the FADEC system is
\be
\ba{l}
\dps x^c_{+} = A^c(\theta) x^c + B^c (\theta) y + B^c_{w}(\theta) w_{-} \cr
\dps u =C^c(\theta) x^c + D^c(\theta) y \cr
\ea
\label{fadecss}
\ee
A diagram of the system in Simulink is given in Figure \ref{fig:fadec}.

For the sake of brevity, we have chosen not to display the closed-form expression of the parameter varying matrices in (\ref{fadecss}) as they are very large. We now seek to compute an ellipsoid invariant for (\ref{fadecss}). There were several issues with the FADEC model that presented a challenge towards finding a single ellipsoid invariant:
\begin{enumerate}
\item There is a sample delay on the input vector $w$. This precluded the usage of a simple quadratic function $V^c(x)=x^{\m{cT}} P^c x^c$ for the stability analysis. Instead,  a Lyapunov-Krasovskii type functional $V^c(x^c,x^c_{-})=x^{\m{cT}} P^c x^c - x_{-}^{\m{cT}} P^c x^c_{-}$ is needed to handle the sample delay. However the resulting LMI cannot be solved by SeDuMi.
\item A convex hull of the system matrices is needed since the system matrices are not linearly parameter-varying.
\item The Butee component contained complex safety limiters in addition to the simple saturation operators.
\end{enumerate}
We discuss each of these issues in the ensuing sections.

\subsubsection{Sample Delay}
The problem with the sample delay was resolved simply by its removal. We shifted some of the dynamics in the controller
forward by one sample and removed the one sample delay on $\hat{u}_{1}$ so that the signal $w$ no longer need to be delayed by one sample. The changes are small enough that the performance of the controller is not noticeably affected as indicated by the simulations. Without the sample delay on $w$, the system in (\ref{fadecss}) becomes the following state-space system
\be
\ba{l}
\dps x^c_{+} = \hat{A}^c(\theta) x^c + \hat{B}^c (\theta) \hat{u} \cr
\dps u =C^c(\theta) x^c + D^c(\theta) y \cr
\ea
\label{fadecss2}
\ee
with a new state-transition matrix $\hat{A}^c$, and
$\hat{B}^c(\theta) =\lb \ba{cc} B^c(\theta) & B^c_{\hat{u}_{1}}(\theta) \ea \rb$, $\dps \hat{u}=\lb \ba{c} y \cr \hat{u}_{1} \ea \rb$.

\subsubsection{Butee Component}
There are two modes of operation to the Butee. The simple operational mode of the Butee is consisted of two identical saturation operator that restricts the input to the Butee and the output from the Butee to the interval $\lb 0.07, 0.098 \rb$. We can model the saturation nonlinearities using a sector-bound inequality.  Let $\delta$ be the midpoint of the output range of the saturation operator i.e. $\delta = 0.5 (0.07 + 0.098)=0.84$. Let the output from the Butee to be denoted by $w$ and let $\tilde{w} = w - \delta$. With $m_{1}=1$ and $0<m_{2}<m_{1}$, we have the sector-bound inequality
\be
\dps (\tilde{y} -  m_{1} C^c x^c - m_{1} D^c u + m_{1} \delta)^{\mathsf{T}} (\tilde{y} -m_{2} C^c x^c - m_{2} D^c u + m_{2} \delta) \leq 0.
\label{sector}
\ee
Let $\kappa_{1}=m_{1} m_{2}$, $\kappa_{2} \dps = \frac{1}{2}(m_{1} + m_{2})$.
The sector bound constraint in (\ref{sector}) is equivalent to the quadratic inequality $\forall x$, $\forall u$, $\forall y$,
\be
\lb \ba{c} x^c \cr u \cr \tilde{y} \cr 1 \ea\rb^{\mathsf{T}}
\lb \ba{cccc}
\kappa_{1} C^{c\mathsf{T}} C^c & \kappa_{1} C^{c\mathsf{T}} D^c & - \kappa_{2} C^{c\mathsf{T}} & \kappa_{2} C^{c\mathsf{T}} \cr
\kappa_{1} D^{c\mathsf{T}} C^c  & \kappa_{1} D^{c\mathsf{T}} D^c & -\kappa_{2} D^{c\mathsf{T}} & \kappa_{2} D^{c\mathsf{T}} \cr
-\kappa_{2} C^c &  -\kappa_{2} D^c & 1 & -1 \cr
\kappa_{2} C^c  & \kappa_{2} D^c   &  -1 & 1\ea \rb
\lb \ba{c} x^c \cr u \cr \tilde{y} \cr 1 \ea\rb \leq 0
\label{sector02}
\ee

In the complex operational mode, the Butee employs a type of min/max switching component typically encountered in engine controllers to handle the performance limits. Although its input-output relations cannot be captured using a sector-bound inequality, however this component is sandwiched between the two saturation operators. We can assume a simple bound on the output of the Butee even when this mode is switched on. In fact, a simple bound on the Butee output $\|\hat{u}_{1}\|\leq 1$ produces better results numerically speaking than using the inequality from (\ref{sector02}).

\subsubsection{Convex Hull of the System Matrices}
With the removal of the sample delays, we have the following linear state-space model of the FADEC (the same as
in (\ref{fadecss2}))
\be
\ba{l}
\dps x^c_{+} = \hat{A}^c(\theta) x^c + \hat{B}^c (\theta) \hat{u} \cr
\dps u =C^c(\theta) x^c + D^c(\theta) y.
\ea
\label{fadecss3}
\ee

\begin{figure}[!ht]
\centering
\subfloat[]{\includegraphics[scale=0.46]{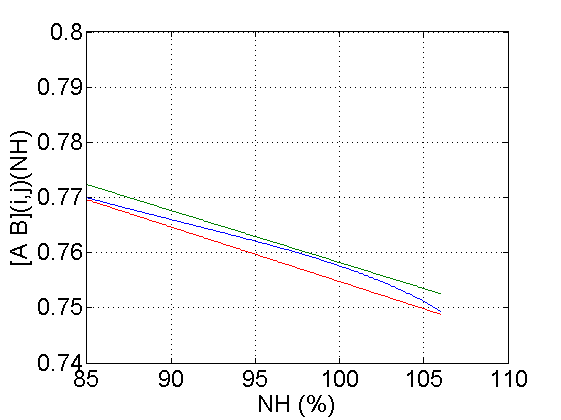}
}
\subfloat[]{\includegraphics[scale=0.46]{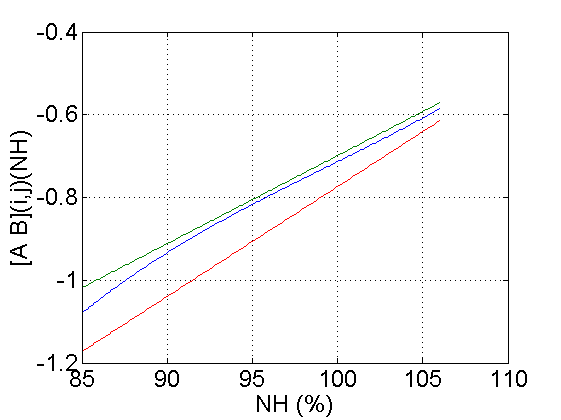}
}
\\
\subfloat[]{\includegraphics[scale=0.46]{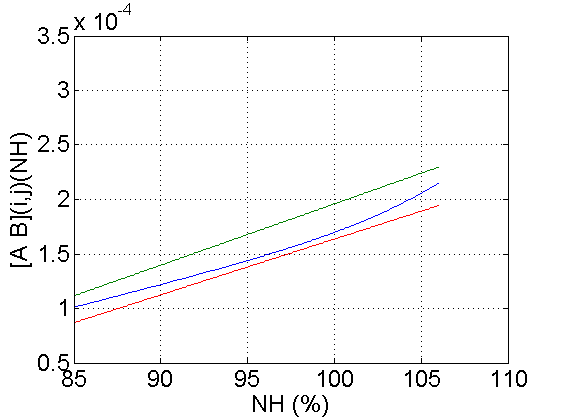}
}
\subfloat[]{\includegraphics[scale=0.46]{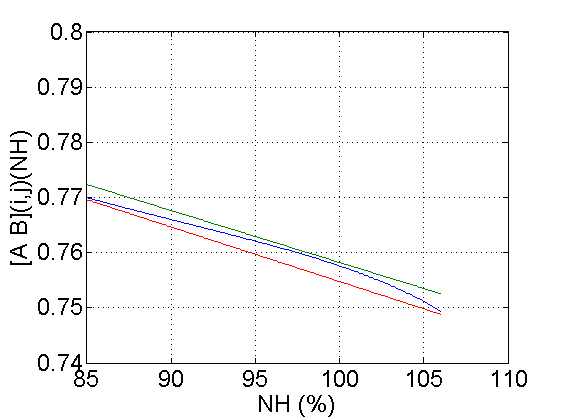}
}
\\
\subfloat[]{\includegraphics[scale=0.46]{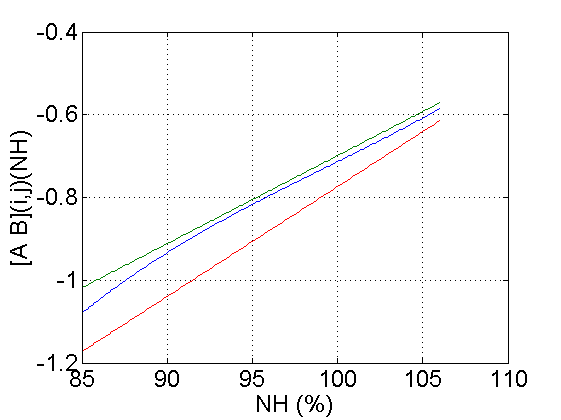}
}
\subfloat[]{\includegraphics[scale=0.46]{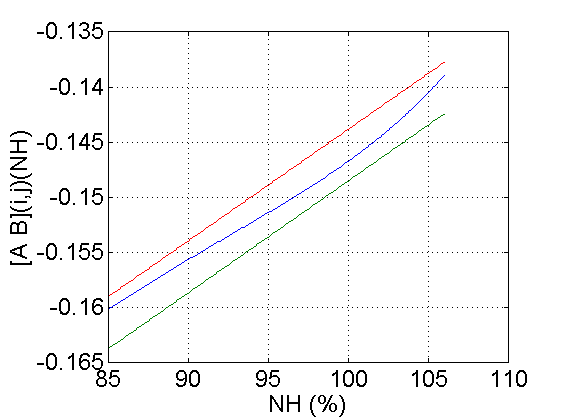}
}
\\
\subfloat[]{\includegraphics[scale=0.46]{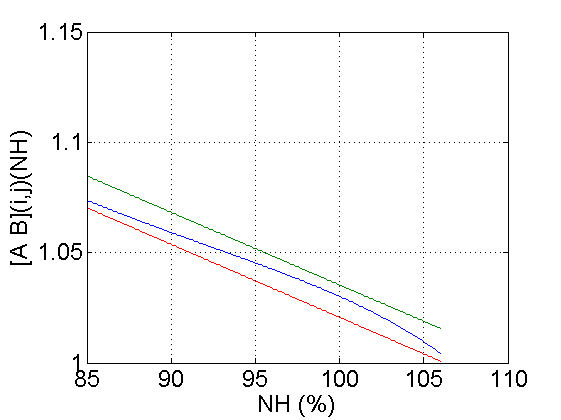}
}
\subfloat[]{\includegraphics[scale=0.46]{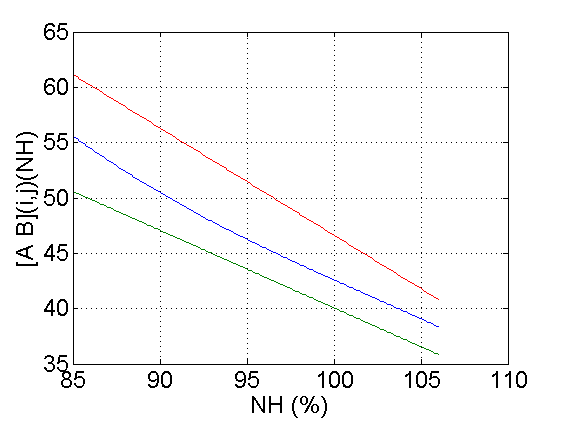}
}
\caption{Varying entries and the convex hull}
\label{plot01}
\end{figure}

To compute a single ellipsoid invariant for (\ref{fadecss3}), we need to find a convex hull that contains all $\hat{A}^c (\theta)$ and $\hat{B}^c (\theta)$ for an assumed range of $\theta$. Since $\theta$ is a polynomial function of the scheduling parameter $\alpha=NH$, the state-transition matrix $\hat{A}$ and the input matrix $\hat{B}^c$ are smooth matrix functions of $NH$. To construct the convex hull, we first compute 4 corner matrices $\lb \ba{cc} \hat{A}^c & \hat{B}^c \ea \rb_{ij},i,j=1,2$ using the following formulas
\be
\ba{c}
\dps \lb \ba{cc} \hat{A}^c & \hat{B}^c \ea \rb_{11}=\lb \ba{cc} \hat{A}^c(\alpha_{\min}) & \hat{B}^c(\alpha_{\min}) \ea \rb + \Delta_{1} \cr
\dps \lb \ba{cc} \hat{A}^c & \hat{B}^c \ea\rb_{12}=\lb \ba{cc} \hat{A}^c(\alpha_{\min}) & \hat{B}^c(\alpha_{\min}) \ea \rb - \Delta_{2} \cr
\dps \lb \ba{cc} \hat{A}^c & \hat{B}^c \ea\rb_{21}=\lb \ba{cc} \hat{A}^c(\alpha_{\max}) & \hat{B}^c(\alpha_{\max}) \ea \rb + \Delta_{3} \cr
\dps \lb \ba{cc} \hat{A}^c & \hat{B}^c \ea\rb_{22}=\lb\ba{cc} \hat{A}^c(\alpha_{\max}) & \hat{B}^c(\alpha_{\max}) \ea \rb - \Delta_{4},
\ea
\label{corner}
\ee
with some estimated perturbation matrices $\Delta_{i},i=1,\ldots,4$. The minimum and maximum values of the scheduling parameter are $\alpha_{\min}=85\%$ and $\alpha_{\max}=106\%$. Next, we need to check if the convex hull of the 4 corner matrices $\Co \lc \lb \ba{cc} \hat{A}^c & \hat{B}^c \ea \rb_{ij}, i,j=1,2\rc$ contains all $\hat{A}^c(\theta)$ and $\hat{B}^c(\theta)$ for the range of $NH$ that we assumed to be valid. A graphical way to check this is to plot each of the entries from $\hat{A}^c$ and $\hat{B}^c$ that varies with NH as a function of NH, and then check if any of the resulting curves violate the convex hull formed by the $4$ corners. If there is a violation in any one entry, then we have to recalculate the 4 corner matrices with larger perturbation matrices until there is no violation in any of the entries.

It turns out that, of the 154 possible entries in the matrices $\hat{A}^c$ and $\hat{B}^c$, only 30 of them vary according to the parameter NH. The rest are either constant or zero. Figure \ref{plot01} has some example plots of those entries as a function of the parameter NH. The blue curves are the plots of the entries of $\hat{A}^c$ or $\hat{B}^c$ that vary with respect to the parameter NH. The red and green lines delineate the convex hull formed by the four corner matrices.
As you can see, all the curves are well within the convex hull formed by the 4 corners.

\subsubsection{Open-Loop Stability Result}
With the 4 corner matrices, we can now apply the following stability criterion to generate an invariant for the system in (\ref{fadecss3}).

\begin{proposition}
Assume the matrix $\lb \ba{cc} \hat{A}^c(\theta) & \hat{B}^c(\theta) \ea \rb \in \Co \lc  \lb \ba{cc} \hat{A}^c & \hat{B}^c \ea \rb_{ij}, i,j=1,2\rc$ for some range of NH and assume that $ \| \hat{u} \| \leq 1$. If there exists a positive definite matrix $P^c$ and a scalar $\xi>0$ that satisfies
\be
\dps \lb \ba{cc}
\hat{A}_{ij}^{c\mathsf{T}} P^c \hat{A}^c_{ij} - P^c + \xi P^c & \hat{A}_{ij}^{c\mathsf{T}} P^c \hat{B}^c_{ij} \\[5pt]
\hat{B}_{ij}^{c\mathsf{T}} P^c \hat{A}^c_{ij} & \hat{B}_{ij}^{c\mathsf{T}} P^c \hat{B}^c_{ij} - \xi I
\ea \rb \prec 0,
\label{lmi05}
\ee
then the set $\lc x^c | x^{c\mathsf{T}} P^c x^c\leq 1\rc$ is an invariant set with respect to (\ref{fadecss3}).
\label{prop06}
\end{proposition}

Using proposition \ref{prop06}, for $\xi=0.02354$, a Lyapunov matrix $P^c \in \Re^{11 \times 11}$ is computed. The numerical value of this matrix is given in Appendix A.

\section{Engine Control Stability Analysis}

\begin{figure}[!ht]
\centering
\includegraphics[width=0.6\textwidth]{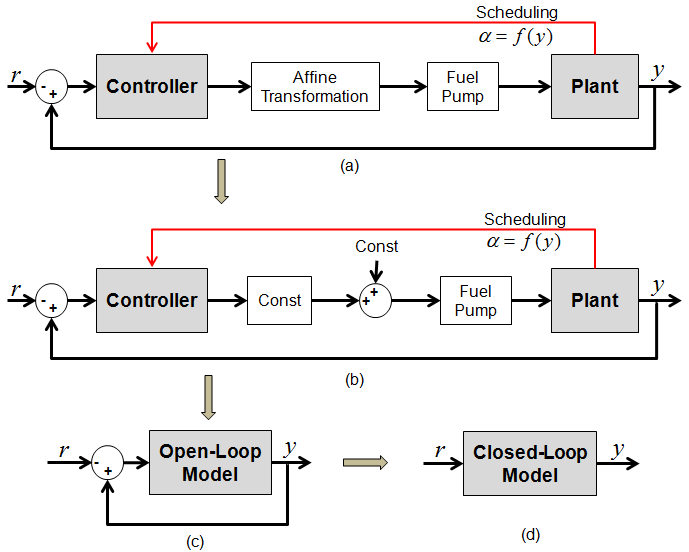}
\caption{Block diagram visualization of the modeling process using plant, controller, and fuel pump models} \label{fig:linear_modeling_process}
\end{figure}

In order to fulfill the stability analysis of the engine with its control system, we needed to obtain the open-loop and closed-loop model of the whole system using engine, controller, and fuel pump dynamics. Figure \ref{fig:linear_modeling_process} visualizes this process. The engine linearization data (including plant, controller, open-loop systems, and closed-loop system matrices) for the four important engine equilibrium points are given in Appendix D.

The states in the linear plant model are the spool speeds and spool accelerations (i.e., $x^p \in \Re^4$), and the input to the plant is the fuel flow rate in kg/hr. The controller has 11 states ($x^c \in \Re^{11}$), and its output is fuel pump RPM in percentage and nondimensional. The output of the controller ($W_c$) is multiplied by $4100/160$ to obtain $W_p=(4100/160) W_c$, then the output signal $W_p$ is transformed using $v=3883 W_p-244.06$; the fuel pump transfer function ($T_{pump}(z)=\frac{W_f(z)}{v(z)}$) is
\begin{equation}
T_{pump}(z)=\frac{0.21756}{z-0.8187},
\end{equation}
where its output, fuel flow rate ($W_f$), is the input to the plant in kg/hr.

Stability analysis of the overall system with controller is done using two different approaches. These methods are the  Bounded Real Lemma \cite{lmi-boyd-1994, NonlinDynSysCont-haddad-2008} and the closed-loop system stability analysis approach developed in \cite{GainSchedStabConf-pakmehr-2013, GSstability-pakmehr-2013}. To obtain a numerical value of stability matrix $P$ for software verification process, there is a set of LMIs in each of these approaches, which can be solved for all $i=\{1,2,...,L\}$ using YALMIP \cite{YALMIP-lofberg-2004} and SeDuMi \cite{sedumi-Sturm-2001} packages in Matlab.

\subsection{Bounded Real Lemma}
Having in mind that $G(z)$ is bounded real if and only if $G(z)$ is asymptotically stable and $||G(z)||_{\infty} \leq \gamma$, the \emph{Bounded Real Lemma} is given as follows
\begin{thm}
Consider the dynamical system
\begin{equation}
\begin{array}{l}
\displaystyle  G_i(z) \overset{\min}{\sim}  \left[
       \begin{array}{c|c}
           A^{ol}_i & B^{ol}_i \\ \hline
           C^{ol}_i & D^{ol}_i
       \end{array}
    \right], ~~~ i=1,2,...,L,
\end{array}
\end{equation}
with input $u(.) \in \mathcal U$ and output $y(.) \in \mathcal Y$. If $\gamma_i I_n-D^{ol^{\mathsf{T}}}_i D^{ol}_i-B^{ol^{\mathsf{T}}}_i P B^{ol^{\mathsf{T}}}_i > 0$ for all $i=\{1,2,...,L\}$, then $G^{ol}_i(z)$ is bounded real if and only if there exist a $P=P^{\mathsf{T}}>0$ for all $i=\{1,2,...,L\}$ such that
\begin{equation}\label{eqn_gs015}
\left[    \begin{array}{cc}
A^{ol^{\mathsf{T}}}_i P A^{ol}-P+C^{ol^{\mathsf{T}}}_i C^{ol}_i & (B^{ol^{\mathsf{T}}}_i P A^{ol}_i+D^{ol^{\mathsf{T}}}_i C^{ol}_i)^{\mathsf{T}} \\[5pt]
(B^{ol^{\mathsf{T}}}_i P A^{ol}_i+D^{ol^{\mathsf{T}}}_i C^{ol}_i) & -(\gamma_i I_n-D^{ol^{\mathsf{T}}}_i D^{ol}_i-B^{ol^{\mathsf{T}}}_i P B^{ol}_i)
\end{array}   \right] \preccurlyeq 0.
\end{equation}
\end{thm}
In this case, the LMI (\ref{eqn_gs015}) is solved for the four main equilibrium points of the system, i.e. $i=4$. A Lyapunov matrix $P \in \Re^{16 \times 16}$ is computed, and the numerical value of this matrix is given in Appendix B.

\subsection{Closed-Loop Stability}
The discussions in this section about the engine closed-loop stability are extended from \cite{GainSchedStabConf-pakmehr-2013, GSstability-pakmehr-2013}.

\begin{thm}\label{thm2}
Consider the closed-loop system
\begin{equation}\label{eqn_gs112}
 x_{+}=F(x,r),
\end{equation}
and assume there is a family of equilibrium points $(x_{eq},r_{eq})$ such that $F(x_{eq},r_{eq})=0$. Define $A^{cl} = \frac{\partial F(.)}{\partial x} \in \overline{S}, ~\forall x \in D_x $, where $\overline{S}$ is the set of linearizations of the system (\ref{eqn_gs112})
\begin{equation}\label{eqn_gs113}
 \overline{S} := \{ A^{cl}, \forall x \in D_x \}.
\end{equation}
Assume there exist symmetric positive definite matrix $P$, such that
\begin{equation}\label{eqn_gs54}
 A^{cl^{\mathsf{T}}} P A^{cl} - P \prec 0, ~~~ \forall A^{cl} \in \overline{S},
\end{equation}
then the system (\ref{eqn_gs112}) is stable. In other words, assuming the initial state is sufficiently close to some equilibrium, then the closed-loop system remains in a neighborhood of the equilibrium manifold for all $t \geq 0$.
\end{thm}

\begin{rmk}\label{rmk1}
In practice we can not obtain $\overline{S}$, instead, we can linearize system (\ref{eqn_gs112}) for a large number of points $x_i$, $i=1, \ldots, L$, which we claim is sufficient to cover the set of actual operating conditions, to show the stability of the closed-loop system. Define $S$ as a matrix polytope described by its vertices
\begin{equation}\label{eqn_gs114}
 S:= \mathrm{Co}\{ A^{cl}_{1}, ..., A^{cl}_{L} \},
\end{equation}
where $A^{cl}_{i} = \left. \frac{\partial F(.)}{\partial x(t)} \right|_{x=x_i} \in S$, for all $i \in \{ 1,2, ..., L \}$. Note that $A^{cl}_{i}$ can be obtained by linearizing the nonlinear system (\ref{eqn_gs112}) at non-equilibrium points (transient condition), and also at equilibrium points (steady state condition). Then using convex optimization tools \cite{YALMIP-lofberg-2004, sedumi-Sturm-2001}, we compute a common symmetric positive definite matrix $P$, such that
\begin{equation}\label{eqn_gs115}
 A^{cl^{\mathsf{T}}}_{i} P A^{cl}_{i} - P \prec 0, ~~~ \forall i \in \{ 1,2, ..., L \}.
\end{equation}
\end{rmk}
In this case, the LMI (\ref{eqn_gs115}) is solved for the four main equilibrium points of the system, i.e. $i=4$. A Lyapunov matrix $P \in \Re^{16 \times 16}$ is computed, and the numerical value of this matrix is given in Appendix C.

\section{Autocoded C with Proof Annotations}

First we give a brief introduction to the formal C annotation language ACSL. A main function of the ACSL is that it can be used to formally specify properties about the variables of the code using an annotation language that is similar in syntax and semantics as the C language. 
The specified properties can be either assumptions or inductive invariants. The former case does not require a proof as it is an assumption made on the variable(s). For example, for a real-time system that interacts with the environment, we need to assume some bounds on inputs from the environment. 
The latter does require a proof on the level of the code. As mentioned in section \ref{autocodeProcess}, a plethora of tools exist that can be used to analyze ACSL annotations and discharge any necessary proof obligations.   
\begin{figure}[!ht] 
\begin{lstlisting}
/*@
	assume input<=1
	ensure x<=1
*/
float x=input; 
/*@ 
	require x*x<=1 
        ensures x*x<=1
*/
while (1) { 
	x=0.99*x;
}
\end{lstlisting}
\caption{Simple ACSL example} 
\label{loop01}
\end{figure}

Typically in ACSL, the assumptions are specified using the keyword \emph{assume}. For example, in the C code shown in Figure \ref{loop01}, the first line of the code assigns the value of the variable $input$ to the variable $x$. We want to assume that the variable $input$ is bounded by $1$ so we inserted an ACSL statement, which is encapsulated within the symbols /*@ and */, that specifies this property. The keywords \emph{require} and \emph{ensures} are used to specify the invariants. The former denotes the valid condition before the execution of the line of the code e.g. the \emph{pre-condition} while the latter denotes the valid condition afterwards e.g. the \emph{post-condition}. In Figure \ref{loop01}, we have the invariant $x*x<=1$ which holds true throughout the execution of the infinite loop. These type of invariants are expressed as both a pre- and a post-condition for the loop. 
\begin{figure}[!ht]
\begin{lstlisting}
/*@ 
        requires in_ellipsoidQ(QMat_0,vect_of_11_scalar(_state_->delay_aw0_memory,_state_->delay_aw1_memory,_state_->delay_E0_memory,_state_->delay_E1_memory,_state_->delay_D0_memory,_state_->delay_D1_memory,_state_->delay_x1_memory,_state_->delay_x2_memory,_state_->delay_aw2_memory,_state_->delay_E2_memory,_state_->delay_D2_memory));
        requires \valid(_io_) && \valid(_state_);
        ensures in_ellipsoidQ(QMat_1,vect_of_11_scalar(_state_->delay_aw0_memory,_state_->delay_aw1_memory,_state_->delay_E0_memory,_state_->delay_E1_memory,_state_->delay_D0_memory,_state_->delay_D1_memory,_state_->delay_x1_memory,_state_->delay_x2_memory,_state_->delay_aw2_memory,_state_->delay_E2_memory,_state_->delay_D2_memory));
*/
void pla_compute(t_pla_io *_io_, t_pla_state *_state_) {
    REAL NL;
    REAL NH;
    REAL P3_KPa_;
    REAL PLA;
\end{lstlisting}
\caption{Ellipsoid invariants for the generated FADEC code}
\label{fadfunc}
\end{figure}

For the credible autocoding of the FADEC, we inserted the open-loop stability proof into the Simulink diagram. The autocoding process generate two functions. 
The first one is the initialization function and the other, called the $pla\rm{\_}compute$, is an amalgamation of the state-transition and output functions of the FADEC. The stability proof is inserted as an ellipsoid invariant on the function $pla\rm{\_}compute$ in two instances. They are specified using both \emph{require} and \emph{require} keywords in the ACSL comment shown in Figure \ref{fadfunc}. The function $in\rm{\_}ellipsoidQ$ defines the ellipsoid invariant using two arguments: the ellipsoid matrix in the form $Q=P^{-1}$ and vector of variables that is captured by the ellipsoid set. 
\begin{figure}[!ht]
\begin{lstlisting}
    /*@ 
            behavior ellipsoid544_1:
            requires in_ellipsoidQ(QMat_562,vect_of_13_scalar(_state_->delay_aw1_memory,_state_->delay_aw2_memory,Sum_of_Elements12_1,Sum_of_Elements12_2,_state_->delay_aw0_memory,_state_->delay_E0_memory,_state_->delay_E1_memory,_state_->delay_D0_memory,_state_->delay_D1_memory,_state_->delay_x1_memory,_state_->delay_x2_memory,_state_->delay_E2_memory,_state_->delay_D2_memory));
            ensures in_ellipsoidQ(QMat_563,vect_of_11_scalar(_state_->delay_aw0_memory,_state_->delay_aw1_memory,_state_->delay_E0_memory,_state_->delay_E1_memory,_state_->delay_D0_memory,_state_->delay_D1_memory,_state_->delay_x1_memory,_state_->delay_x2_memory,_state_->delay_aw2_memory,_state_->delay_E2_memory,_state_->delay_D2_memory));
            @ PROOF_TACTIC (use_strategy (AffineEllipsoid));
    */
 {
    _state_->delay_aw2_memory = Sum_of_Elements12_2;
 }
}

\end{lstlisting}
\caption{Verification of the invariant using the generated post-condition} 
\label{funcend}
\end{figure}

The reason for specifying the ellipsoid invariant twice, as pre and post-conditions, is because the function $pla\rm{\_}compute$ is executed in a loop just like the loop shown in Figure \ref{loop01}. Any invariant that is valid before the execution of $pla\rm{\_}compute$ also need to be valid after its execution. 
This is a property that needs to be proven on the C code level and the backend tools mentioned in section \ref{autocodeProcess} have been equipped to handle this type of  proof obligation. To enable the backend analyzer, the credible autocoder also generates additional ellipsoid invariants along with the proof strategy used for every line of code inside the function $pla\rm{\_}compute$. This is done until the last line of the function as shown in Figure \ref{funcend}, in which there is a post-condition generated by the autocoder This generated post-condition is used to check against the ellipsoid invariant that was inserted as pre- and post-conditions on the function $pla\rm{\_}compute$. For proof of correctness, one just need to show that the latter implies the former. 

\section{Simulation Results}

\begin{figure}[!ht]
\centering
\subfloat[]{\includegraphics[scale=0.48]{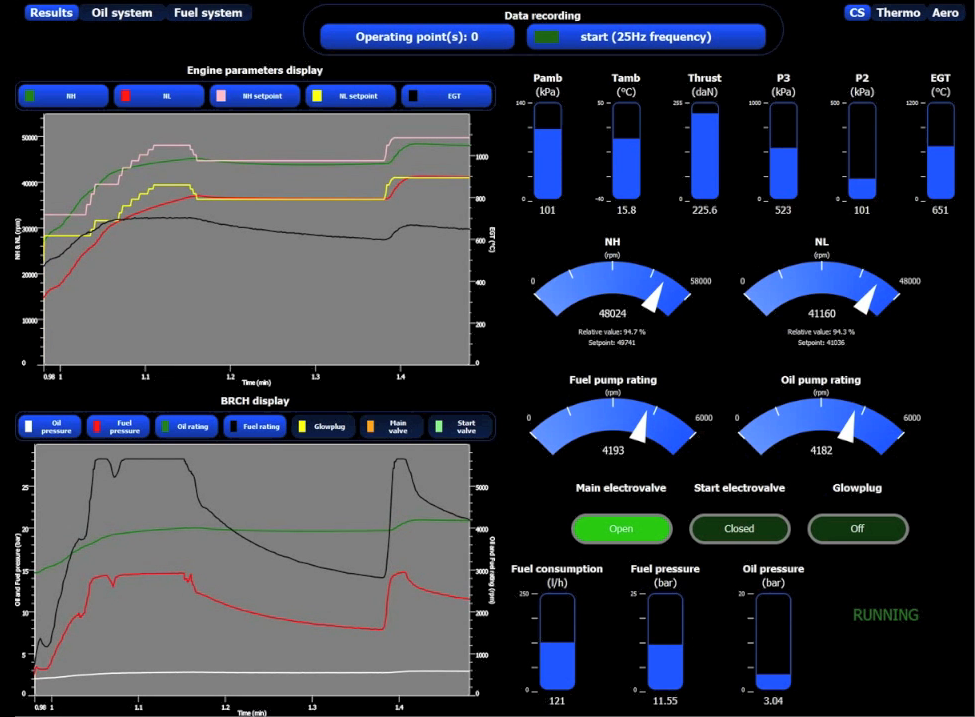} } \\
\subfloat[]{\includegraphics[scale=0.48]{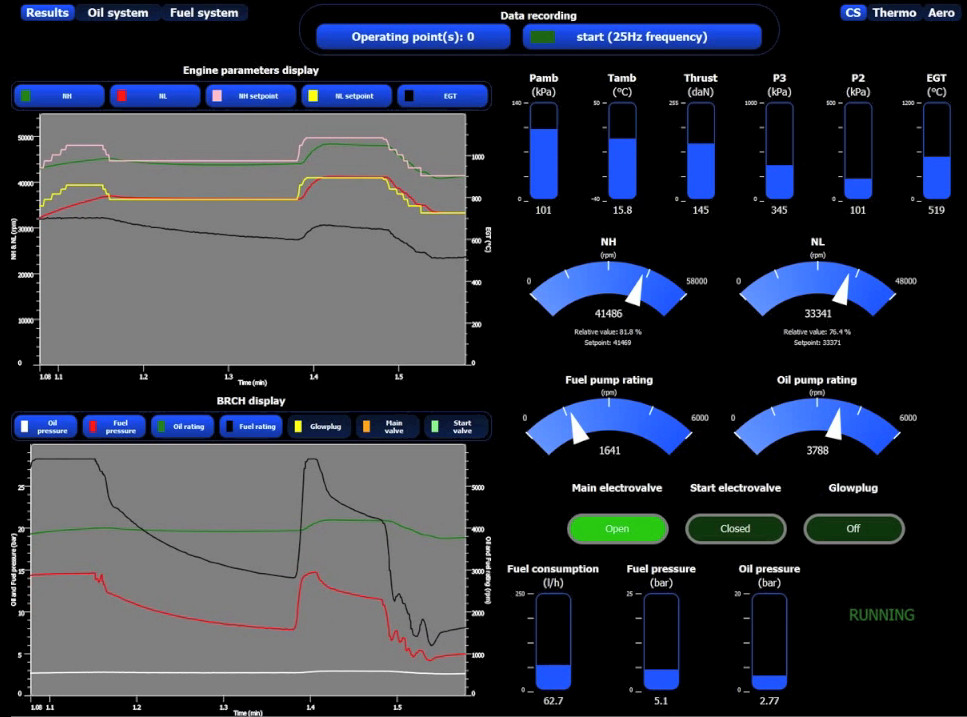} }
\caption{Snapshots of the verifiable engine controller implementation on the DGEN 380 turbofan engine virtual test bench}
\label{fig:simulation_engine}
\end{figure}

Here we present the simulation results to show that the commented code works like the uncommented one. The code verification process is developed to be sure that the controller is certifiable; it is not supposed to replace the simulation process. We transform the proof into an observer to detect the system's malfunctions and to indicate it is actually a \textit{health monitoring} software.

Figure \ref{fig:simulation_engine} shows two snapshots of the WESTT command screen for the case where the verifiable controller implemented on the DGEN 380 turbofan engine virtual test bench. These snapshots illustrate evolution of both engine spool speeds that closely follow their reference signals, and also engine fuel and oil pressure time histories.

A visualizations of the engine related avionics is also presented in these snapshots. The snapshot shows real-time measurement of pressures ($P_{amb}$, $P_2$, and $P_3$), temperatures ($T_{amb}$, and $EGT$), speeds ($NH$, and $NL$), thrust, fuel pump rating, oil pump rating, fuel consumption, fuel pressure, and oil pressure.

\section{Conclusions}
A stability proof was presented for the closed-loop DGEN 380 turbofan engine system with its gain scheduled controller based on the Lyapunov stability theory. Using convex optimization tools, various numerical values of the Lyapunov stability matrix, for the closed-loop system and the controller, were computed separately. With these stability matrices, ellipsoid invariant sets were constructed, which were used efficiently for DGEN turbofan engine control code stability analysis and verification. The verifiable engine controller code then implemented successfully on the engine virtual test bench (WESTT) to control a high fidelity DGEN 380 engine model. Simulation results are presented to illustrate the efficiency of the presented approach. The code verification framework presented here, hopefully, can be used for gas turbine engine control software certification in the future.

\section*{Acknowledgment}
This material is based upon the work supported by the Air Force Office of Scientific Research (AFOSR), the National Aeronautics and Space Administration (NASA), and the National Science Foundation (NSF).

\bibliographystyle{plain}

\appendix       

\section*{Appendix A: Numerical Value of the Controller Lyapunov Matrix $P^c$} 
{\scriptsize
\be
\ba{l}
\dps P^c=\lb\ba{ccccccccccc}
  0.3214& -0.0374&  0.0940& -0.0103&  0.0049& -0.0000&-0.0000& -0.0005& -0.0798& -0.0324& -0.0039\cr
 -0.0374&  0.0494&  0.0176&  0.0127&  0.0008& -0.0000&0.0000& -0.0004& -0.0065& -0.0087& -0.0010\cr
  0.0940&  0.0176&  0.1565&  0.0031&  0.0086& -0.0000&-0.0000&  0.0000& -0.0402& -0.0546& -0.0080\cr
 -0.0103&  0.0127&  0.0031&  0.0098&  0.0007& -0.0000&0.0000&  0.0007& -0.0014& -0.0021& -0.0006\cr
  0.0049&  0.0008&  0.0086&  0.0007&  0.0105&  0.0000&-0.0000& -0.0000& -0.0048& -0.0078& -0.0063\cr
 -0.0000& -0.0000& -0.0000& -0.0000&  0.0000&  0.0491&0.0000&  0.0000&  0.0000&  0.0000& -0.0000\cr
 -0.0000&  0.0000& -0.0000&  0.0000& -0.0000&  0.0000&0.0491&  0.0002&  0.0000&  0.0000&  0.0000\cr
 -0.0005& -0.0004&  0.0000&  0.0007& -0.0000&  0.0000&0.0002&  0.0501& -0.0001& -0.0001& -0.0000\cr
 -0.0798& -0.0065& -0.0402& -0.0014& -0.0048&  0.0000&0.0000& -0.0001&  0.2917&  0.1077&  0.0055\cr
 -0.0324& -0.0087& -0.0546& -0.0021& -0.0078&  0.0000&0.0000& -0.0001&  0.1077&  0.1500&  0.0084\cr
 -0.0039& -0.0010& -0.0080& -0.0006& -0.0063& -0.0000&0.0000& -0.0000&  0.0055&  0.0084&  0.0104
 \ea \rb
 \ea\\
\label{p03}
\ee
}

\section*{Appendix B: Numerical Value of the System Lyapunov Matrix $P$ Computed using Bounded Real Lemma} 
{\scriptsize
\be
\ba{l}
\dps P=1.0e+011 \\
\lb\ba{cccccccc}
    0.3011  &  0.3056 & -0.3304 & -0.3355 & -0.0000 &  0.0077 &  0.0005 & 0.0000 \cr
    0.3056  &  0.3103 & -0.3354 & -0.3406 & -0.0000 &  0.0079 &  0.0005 & 0.0000 \cr
   -0.3304  & -0.3354 &  0.3765 &  0.3823 &  0.0000 & -0.0103 & -0.0006 &-0.0000 \cr
   -0.3355  & -0.3406 &  0.3823 &  0.3882 &  0.0000 & -0.0104 & -0.0006 &-0.0000 \cr
   -0.0000  & -0.0000 &  0.0000 &  0.0000 &  0.0000 & -0.0000 & -0.0000 &-0.0000 \cr
    0.0077  &  0.0079 & -0.0103 & -0.0104 & -0.0000 &  1.2370 & -0.0128 & 0.0019 \cr
    0.0005  &  0.0005 & -0.0006 & -0.0006 & -0.0000 & -0.0128 &  0.0419 & 0.0001 \cr
    0.0000  &  0.0000 & -0.0000 & -0.0000 & -0.0000 &  0.0019 &  0.0001 & 0.0000 \cr
   -0.0001  & -0.0001 &  0.0002 &  0.0002 &  0.0000 & -0.0046 &  0.0098 & 0.0000 \cr
    0.0009  &  0.0009 & -0.0011 & -0.0012 & -0.0000 &  0.0203 &  0.0008 & 0.0001 \cr
    0.0000  &  0.0000 & -0.0000 & -0.0000 & -0.0000 & -0.0000 & -0.0000 &-0.0000 \cr
    0.0000  &  0.0000 & -0.0000 & -0.0001 & -0.0000 &  0.0001 &  0.0001 & 0.0000 \cr
    0.0046  &  0.0047 & -0.0058 & -0.0060 & -0.0000 &  0.0236 &  0.0088 & 0.0002 \cr
    0.0028  &  0.0029 & -0.0022 & -0.0024 & -0.0000 & -0.4039 & -0.0139 &-0.0004 \cr
   -0.0000  & -0.0000 &  0.0000 &  0.0000 &  0.0000 & -0.0018 & -0.0001 &-0.0000 \cr
   -0.0010  & -0.0010 &  0.0013 &  0.0013 &  0.0000 & -0.0176 & -0.0007 &-0.0001
 \ea  \right. \\[10pt]
 \left. \ba{cccccccc}
     -0.0001 & 0.0009 & 0.0000 & 0.0000 &  0.0046  &  0.0028  & -0.0000 & -0.0010 \cr
     -0.0001 & 0.0009 & 0.0000 & 0.0000 &  0.0047  &  0.0029  & -0.0000 & -0.0010 \cr
	  0.0002 &-0.0011 &-0.0000 &-0.0000 & -0.0058  & -0.0022  &  0.0000 &  0.0013 \cr
	  0.0002 &-0.0012 &-0.0000 &-0.0001 & -0.0060  & -0.0024  &  0.0000 &  0.0013 \cr
	  0.0000 &-0.0000 &-0.0000 &-0.0000 & -0.0000  & -0.0000  &  0.0000 &  0.0000 \cr
     -0.0046 & 0.0203 &-0.0000 & 0.0001 &  0.0236  & -0.4039  & -0.0018 & -0.0176 \cr
	 -0.0046 & 0.0203 &-0.0000 & 0.0001 &  0.0236  & -0.4039  & -0.0018 & -0.0176 \cr
	  0.0000 & 0.0001 &-0.0000 & 0.0000 &  0.0002  & -0.0004  & -0.0000 & -0.0001 \cr
	  0.0059 & 0.0005 &-0.0000 & 0.0000 &  0.0057  & -0.0056  & -0.0000 & -0.0004 \cr
	  0.0005 & 0.0599 & 0.0000 &-0.0001 & -0.0025  &  -0.0109 & -0.0001 & -0.0573 \cr
	 -0.0000 & 0.0000 & 0.1849 & 0.0000 &  0.0000  &  0.0000  &  0.0000 & -0.0000 \cr
	  0.0000 &-0.0001 & 0.0000 & 0.1848 &  0.0016  &  0.0016  & -0.0000 &  0.0001 \cr
      0.0057 &-0.0025 & 0.0000 & 0.0016 &  0.2753  &  0.0579  & -0.0002 &  0.0028 \cr
	 -0.0056 &-0.0109 & 0.0000 & 0.0016 &  0.0579  &  0.3842  &  0.0002 &  0.0136 \cr
	 -0.0000 &-0.0001 & 0.0000 &-0.0000 & -0.0002  &  0.0002  &  0.0000 &  0.0001 \cr
	 -0.0004 &-0.0573 &-0.0000 & 0.0001 &  0.0028  & 0.0136   &  0.0001 &  0.0602
 \ea \rb
 \ea
\label{p01}
\ee
}
where its condition number is 7.7519e+011.

\section*{Appendix C: Numerical Value of the System Lyapunov Matrix $P$ Computed using Closed-Loop Stability Approach} 
{\scriptsize
\be
\ba{l}
\dps P= \\
\lb\ba{cccccccc}
   11.4320 & 11.4839  & -1.1424 & -1.1542 & -0.0000 & -0.0107 & -0.0014 & 0.0000 \\
   11.4839 &  11.5361 & -1.1582 & -1.1702 & -0.0000 & -0.0107 &  0.0004 & 0.0000 \\
   -1.1424 &  -1.1582 &  1.8426 &  1.8627 & -0.0000 &  0.0409 &  0.0156 & 0.0001 \\
   -1.1542 &  -1.1702 &  1.8627 &  1.8832 & -0.0000 &  0.0410 &  0.0136 & 0.0001 \\
   -0.0000 &  -0.0000 & -0.0000 & -0.0000 &  0.0000 & -0.0000 & -0.0000 &-0.0000 \\
   -0.0107 &  -0.0107 &  0.0409 &  0.0410 & -0.0000 &  5.1277 & -0.2275 & 0.0070 \\
   -0.0014 &   0.0004 &  0.0156 &  0.0136 & -0.0000 & -0.2275 &  1.5941 & 0.0053 \\
    0.0000 &   0.0000 &  0.0001 &  0.0001 & -0.0000 &  0.0070 &  0.0053 & 0.0001 \\
    0.0005 &   0.0005 &  0.0003 &  0.0003 & -0.0000 & -0.2891 &  0.5244 & 0.0009 \\
    0.0011 &   0.0011 & -0.0005 & -0.0005 &  0.0000 &  0.0730 &  0.0065 & 0.0003 \\
   -0.0000 &  -0.0000 &  0.0000 &  0.0000 & -0.0000 & -0.0006 &  0.0001 &-0.0000 \\
    0.0001 &   0.0001 & -0.0000 & -0.0000 & -0.0000 &  0.0000 & -0.0007 &-0.0000 \\
    0.0074 &   0.0084 &  0.0045 &  0.0029 & -0.0000 &  0.1018 &  0.4834 & 0.0023 \\
    0.0489 &   0.0501 & -0.0354 & -0.0366 & -0.0000 & -4.6693 & -0.0520 &-0.0063 \\
    0.0000 &  -0.0000 & -0.0001 & -0.0001 &  0.0000 & -0.0070 & -0.0053 &-0.0001 \\
   -0.0011 &  -0.0010 &  0.0005 &  0.0005 & -0.0000 & -0.0730 & -0.0067 &-0.0003
 \ea  \right. \\[10pt]
 \left. \ba{cccccccc}
    0.0005 & 0.0011 & -0.0000 &  0.0001 &  0.0074 &  0.0489 &  0.0000 & -0.0011 \\
	0.0005 & 0.0011 & -0.0000 &  0.0001 &  0.0084 &  0.0501 & -0.0000 & -0.0010 \\
	0.0003 &-0.0005 &  0.0000 & -0.0000 &  0.0045 & -0.0354 & -0.0001 &  0.0005 \\
	0.0003 &-0.0005 &  0.0000 & -0.0000 &  0.0029 & -0.0366 & -0.0001 &  0.0005 \\
   -0.0000 & 0.0000 & -0.0000 & -0.0000 & -0.0000 & -0.0000 &  0.0000 & -0.0000 \\
   -0.2891 & 0.0730 & -0.0006 &  0.0000 &  0.1018 & -4.6693 & -0.0070 & -0.0730 \\
	0.5244 & 0.0065 &  0.0001 & -0.0007 &  0.4834 & -0.0520 & -0.0053 & -0.0067 \\
	0.0009 & 0.0003 & -0.0000 & -0.0000 &  0.0023 & -0.0063 & -0.0001 & -0.0003 \\
	0.4449 & 0.0412 &  0.0000 & -0.0003 &  0.0037 & -0.2336 & -0.0009 & -0.0409 \\
	0.0412 & 0.2870 &  0.0000 &  0.0000 & -0.0169 & -0.1316 & -0.0003 & -0.2869 \\
	0.0000 & 0.0000 &  1.0019 &  0.0000 & -0.0000 &  0.0005 &  0.0000 & -0.0000 \\
   -0.0003 & 0.0000 &  0.0000 &  1.0012 & -0.0004 &  0.0003 &  0.0000 & -0.0000 \\
	0.0037 &-0.0169 & -0.0000 & -0.0004 &  0.2953 &  0.1384 & -0.0023 &  0.0168 \\
   -0.2336 &-0.1316 &  0.0005 &  0.0003 &  0.1384 &  5.0994 &  0.0063 &  0.1317 \\
   -0.0009 &-0.0003 &  0.0000 &  0.0000 & -0.0023 &  0.0063 &  0.0001 &  0.0003 \\
   -0.0409 &-0.2869 & -0.0000 & -0.0000 &  0.0168 &  0.1317 &  0.0003 &  0.2869
 \ea \rb
 \ea
\label{p02}
\ee
}
where its condition number is 5.4760e+012.

\section*{Appendix D: Numerical Value of the Matrices for the Engine Equilibrium Points} 
The numerical values for important equilibrium conditions of the DGEN 380 turbofan engine are given in this Appendix. These equilibrium conditions are idle, maximum recommended cruise (MCR), maximum continuous climb (MCM), and take-off power (TOP). The reference values for spool speeds are $NH_{ref}=507.19$ and $NL_{ref}=436.36$.\\

\noindent \textbf{1-Idle}\\
Equilibrium values:
$NH_{eq}=39315~(RPM)$, $NL_{eq}=26181~(RPM)$, $Wf_{eq}=43.67~(kg/hr)$, $PLA_{eq}=0~(deg)$. The nondimensional values of the main plant states in the code are $NH_{idle}=\frac{NH_{eq}}{NH_{ref}}=77.5153 \%$, and $NL_{idle}=\frac{NH_{eq}}{NH_{ref}}=59.9986 \%$. The other states in the plant model are the spool accelerations.\\

Plant matrices
{\footnotesize
\begin{eqnarray}
\begin{array}{l}
A^p_1 = \left[    \begin{array}{cccc}
    1.9746  & 1.0000  &     0    &     0\\
   -0.9747  &    0    &     0    &     0\\
         0  &    0    & 1.9742   & 1.0000\\
         0  &    0    & -0.9742  &      0
\end{array}   \right], ~~
B^p_1 = \left[    \begin{array}{c}
      0.0089\\
     -0.0089\\
      0.0083\\
     -0.0082
\end{array}   \right], ~
C^p_1 = \left[    \begin{array}{cccc}
      1  &   0  &   0  &   0 \\
      0  &   0  &   1  &   0 \\
      0  &   0  &   0  &   0 \\
      0  &   0  &   0  &   0
\end{array}   \right], ~
D^p_1 = 0.
\end{array}
\end{eqnarray}
}
Controller matrices
{\scriptsize
\begin{equation}
\begin{array}{l}
A^c_1   = \left[    \begin{array}{cccccccccccc}
      1.0000 &      0 &   0.0001  &       0  &       0 &        0 &        0 &     0   &    0    &     0     &    0 \\
         0   & 1.0000 &        0  &  0.3000  &       0 &        0 &        0 &     0   &    0    &     0     &    0 \\
  -65.4476   &      0 &   0.7765  &       0  & -1.3768 &        0 &        0 &     0   &    0    &     0     &    0\\
    1.2500   &-1.2500 &   0.0001  & -0.3750  &       0 &  -0.0040 &        0 &     0   &    0    &     0     &    0\\
   -1.0991   &      0 &  -0.0038  &       0  & -0.1682 &        0 &        0 &      0  &     0   &      0    &     0\\
         0   &      0 &        0  &       0  &       0 &  -0.0566 &        0 &      0  &     0   &      0    &     0\\
         0   &      0 &        0  &       0  &       0 &        0 &  -0.0476 &      0  &     0   &      0    &     0\\
         0   & 0.0168 &        0  &  0.0050  &       0 &        0 &  -0.0168 &  -0.1667 &    0   &      0    &     0\\
         0   &      0 &        0  &       0  &       0 &        0 &        0 &      0   & 1.0000  &  0.0001  &   0\\
         0   &      0 &        0  &       0  &       0 &        0 &        0 &      0   & -65.4476 &  0.7765 &  -1.3768\\
         0   &      0 &        0  &       0  &       0 &        0 &        0 &      0   & -1.0991  & -0.0038 &  -0.1682
\end{array}   \right], \\[5pt]
B^c_1 = \left[    \begin{array}{ccc}
         0    &     0    &     0  \\
         0    &     0    &     0\\
   65.4476    &     0    &     0\\
         0    &     0   & -1.2500\\
    1.0991    &     0    &     0\\
         0    &     0    &     0\\
         0    &     0    &     0\\
         0    & 0.0134   & 0.0168\\
         0    &     0    &     0\\
   65.4476    &     0    &     0\\
    1.0991    &     0    &     0
\end{array}   \right], ~~
C^c_1 = \left[    \begin{array}{c}
         0  \\
         0.0174  \\
         0    \\
         0.0052 \\
         0  \\
         0  \\
         -0.0173   \\
         -0.9722  \\
         1.0000    \\
         0.0001 \\
          0
\end{array}   \right]^\mathsf{T}, ~
D^c_1 =  \left[ 0 ~  0.0139  ~  0.0174 \right].
\end{array}
\end{equation}
}
Open-loop matrices
{\tiny
\begin{eqnarray}
\begin{array}{l}
A^{ol}_1 = 1.0e+004 \\
\left[    \begin{array}{cccccccccccccccc}
   0.0002& 0.0001&   0  &   0  &   0.0000 &  0  &  0&   0 &   0 &   0&    0&    0 &     0 &    0 &    0  &    0 \\
  -0.0001&   0   &   0  &   0  &  -0.0000 & 0  & 0   & 0 &  0  &   0  &   0  &   0&     0&   0 &    0  &     0 \\
      0  &   0   &0.0002&  0.0001&  0.0000&  0  &  0  &  0 &  0  &  0  &  0  &   0  &   0  &   0  &   0  &    0\\
      0  &   0  &-0.0001&   0  &-0.0000 &  0  &  0  &  0  & 0 &  0  &  0  &   0   &   0 &    0&     0 &     0 \\
      0  &   0  &   0  &  0  &  0.0001&  0 &  0.1727 & 0 & 0.0518 & 0 &    0 & -0.1723 &-9.6753&  9.9517 & 0.0006 & 0 \\
      0  &   0  &   0  &   0  &   0 &  0.0001 & 0&  0.0000 &0 &   0 &   0 &    0  &   0 &    0 &    0 &    0 \\
      0  &   0  &   0  &   0  &   0 &     0 & 0.0001 & 0 & 0.0000 & 0&    0 &    0 &    0 &    0 &   0 &    0 \\
      0  &   0  &   0  &   0  &   0 & -0.0065 & 0 & 0.0001 & 0 & -0.0001 & 0 &   0 &   0 &    0 &   0  &    0 \\
      0  &   0  &   0  &   0  &  0  &0.0001 &-0.0001 & 0.0000& -0.0000&  0  & -0.0000&   0  &   0 &  0 &  0 &  0\\
      0  &   0  &   0  &   0  &     0 & -0.0001 &  0 &  -0.0000 & 0 &  -0.0000  &   0   &    0  &   0 &  0  &  0 &  0\\
      0  &   0  &   0  &   0  &    0  &    0  & 0  &   0 &   0 &  0  & -0.0000 &   0   &   0  & 0  &  0  &   0 \\
      0  &   0  &   0   &   0  &   0   &  0  &  0 &    0 &    0  &   0   &   0 &  -0.0000  &  0  &  0  &  0  &  0\\
      0  &   0  &  0    &  0   &  0    &0 &  0.0000  &  0  & 0.0000  & 0 &  0 & -0.0000 & -0.0000  & 0 &  0  &  0 \\
      0  &   0  &   0   &  0   &   0   &  0  &   0   &  0  &  0  &   0   &  0   &  0  &  0 & 0.0001 &  0.0000  &  0 \\
      0  &  0   &   0   &  0   &   0   &  0  &  0   & 0   &  0   &  0   &  0  &   0  & 0 & -0.0065 &  0.0001 &-0.0001\\
      0  &  0   &  0    & 0    &  0    &  0  &  0  &   0  &  0   &   0  &   0  &   0  &  0 & -0.0001 & -0.0000 & -0.0000
\end{array}   \right], \\[5pt]
B^{ol}_1 = 1.0e+003 \left[    \begin{array}{cccc}
         0   &      0  &       0  &       0\\
         0   &      0  &       0  &       0\\
         0   &      0  &       0  &       0\\
         0   &      0  &       0  &       0\\
         0   & 1.3814  &  1.7268  & -0.2441\\
         0   &      0  &       0  &       0\\
         0   &      0  &       0  &       0\\
    0.0654   &      0  &       0  &       0\\
         0   &      0  & -0.0013  &       0\\
    0.0011   &      0  &       0  &       0\\
         0   &      0  &       0  &       0\\
         0   &      0  &       0  &       0\\
         0   & 0.0000  &  0.0000  &       0\\
         0   &      0  &       0  &       0\\
    0.0654   &      0  &       0  &       0\\
    0.0011   &      0  &       0  &       0
\end{array}   \right], ~~
C^{ol}_1 = \left[    \begin{array}{cccccccccccccccc}
     1  &   0   &  0  &  0  &  0  &  0  &  0  &  0  &  0  &  0  &  0  &  0  &  0  &   0  &  0  &   0\\
     0  &   0   &  1  &  0  &  0  &  0  &  0  &  0  &  0  &  0  &  0  &  0  &  0  &   0  &  0  &   0\\
     0  &   0   &  0  &  0  &  0  &  0  &  0  &  0  &  0  &  0  &  0  &  0  &  0  &   0  &  0  &   0\\
     0  &   0   &  0  &  0  &  0  &  0  &  0  &  0  &  0  &  0  &  0  &  0  &  0  &   0  &  0  &   0
\end{array}   \right], ~~ D^{ol}_1 = 0, \\[5pt]
||G^{ol}_1(z)||_{\infty} < \gamma_1, ~ \gamma_1 =7.6489e+004.
\end{array}
\end{eqnarray}
}
Closed-loop matrix
{\tiny
\begin{eqnarray}
\begin{array}{l}
A^{cl}_1 = 1.0e+004 \\
\left[    \begin{array}{cccccccccccccccc}
   0.0002 & 0.0001 &   0 &   0  &  0.0000 &   0  &  0  & 0  &   0  &  0 &   0  &   0  &0   &  0  &   0  &  0 \\
  -0.0001 &    0   &   0  &   0 & -0.0000  &  0  &   0  &  0 &   0 &  0 &  0  &  0  & 0  &  0 &   0  &  0 \\
   0      & 0  & 0.0002 &   0.0001  &  0.0000  &  0  &   0  &   0   &   0  &  0  &  0  &  0  &  0  &  0  &   0  &  0 \\
   0     &0 & -0.0001  &  0 &  -0.0000  &  0  &  0  &  0  &   0   &  0  &  0  &   0  &  0   &  0  &  0  &  0 \\
   0     & 0 &  -0.1381 &  0  & 0.0001  & 0  & 0.1727 & 0 & 0.0518  & 0  & 0 & -0.1723 & -9.6753 & 9.9517 & 0.0006 & 0 \\
   0     & 0  &   0  &   0  &   0  & 0.0001  & 0 & 0.0000 &  0  &  0  & 0  &  0  &  0  & 0  &  0  &  0 \\
   0     & 0   &    0  &   0   &  0   &  0  & 0.0001 &   0 & 0.0000 &  0 & 0  &   0  &  0  &  0  &  0  &  0 \\
 -0.0065  & 0  &   0  &   0  &   0 & -0.0065 &  0 &  0.0001 &  0 & -0.0001 &   0 &   0  &  0 &   0 &   0 &  0 \\
   0    & 0    & 0   &  0  &   0  & 0.0001 & -0.0001 & 0.0000 & -0.0000 &  0  &-0.0000 & 0 &  0  & 0  &  0  &  0 \\
 -0.0001  &  0  &   0  &  0 &  0 & -0.0001 & 0 & -0.0000 & 0 & -0.0000 & 0 & 0 &  0 &  0  &  0 &   0 \\
   0    & 0    & 0  & 0  &   0  &  0  & 0  &   0 & 0 &   0 & -0.0000 &   0 &  0  &  0 &   0  & 0 \\
   0    &  0  & 0  &  0  &  0 &  0  & 0 &  0  &  0  &  0  & 0 & -0.0000 &  0  &  0 &  0  &   0 \\
   0   & 0 &  -0.0000   & 0 &   0  & 0  0.0000  &  0 & 0.0000 &   0  &  0  & -0.0000  & -0.0000 &  0 &   0  &  0\\
   0   & 0 &   0 & 0 &  0  &  0 & 0  &  0 &   0  &  0  &  0  & 0  &  0 &  0.0001  & 0.0000  & 0 \\
 -0.0065  & 0 &   0 & 0  &  0  & 0  & 0 &  0 & 0  & 0 &  0  & 0 & 0 & -0.0065  & 0.0001 & -0.0001 \\
 -0.0001  & 0 &  0 & 0  &  0 &  0  & 0  &  0 & 0  &  0  & 0  &  0 &   0 & -0.0001 & -0.0000 & -0.0000
\end{array}   \right].
\end{array}
\end{eqnarray}
}

\textbf{2-MCR (Maximum Recommended Cruise)}\\
Equilibrium values:
$NH_{eq}=48582~(RPM)$, $NL_{eq}=38536~(RPM)$, $Wf_{eq}=86.9~(kg/hr)$, $PLA_{eq}=20~(deg)$. The nondimensional values of the main plant states in the code are $NH_{cr}=\frac{NH_{eq}}{NH_{ref}}=95.7866 \%$, and $NL_{cr}=\frac{NH_{eq}}{NH_{ref}}=88.3124 \%$. \\

Plant matrices
{\footnotesize
\begin{eqnarray}
\begin{array}{l}
A^p_2 = \left[    \begin{array}{cccc}
    1.9746  & 1.0000  &     0    &     0\\
   -0.9747  &    0    &     0    &     0\\
         0  &    0    & 1.9742   & 1.0000\\
         0  &    0    & -0.9742  &      0
\end{array}   \right], ~~
B^p_2 = \left[    \begin{array}{c}
      0.0056\\
     -0.0055\\
      0.0063\\
     -0.0063
\end{array}   \right], ~
C^p_2 = C^p_1, ~ D^p_2 = D^p_1.
\end{array}
\end{eqnarray}
}
Controller matrices
{\scriptsize
\begin{eqnarray}
\begin{array}{l}
A^c_2   = \left[    \begin{array}{cccccccccccc}
    1.0000  &   0  &  0.0001  &   0  &  0  &  0  &   0  &  0  & 0  &  0  &  0\\
       0  &  1.0000  &  0  &  0.3000  &  0 &  0  & 0  & 0 &  0  &   0  &  0\\
  -45.6448  &  0  &  0.7614 & 0  & -0.8004  &  0  &   0 & 0  & 0  &  0  & 0\\
    1.2500 &  -1.2500 &   0.0001 &  -0.3750  &  0 &  -0.0040 &  0  &  0 &  0  & 0 &    0\\
   -1.0431 &  0 &  -0.0055 &   0 &  -0.1507 &  0 &   0  &   0 &  0 &   0 &    0\\
       0  &  0   &   0  &   0 &   0 &  -0.0566  &  0 &   0 &  0 &   0  &  0\\
       0  &   0  &   0  &   0  &   0   & 0  & -0.0476  & 0  & 0  &  0 &   0\\
       0  &  0.0229 & 0  &  0.0069 & 0 &  0 &  -0.0228 &  -0.1667 &  0  & 0 &  0\\
       0  &   0 &    0  &  0  &   0  & 0  &   0  & 0 & 1.0000 &  0.0001 &  0\\
       0  &   0  &   0  & 0  &  0  &  0  &  0  &  0 & -45.6448 & 0.7614 & -0.8004\\
       0   &  0  &   0  &  0  &   0  &  0  &  0  &  0 & -1.0431 &  -0.0055 &  -0.1507
\end{array}   \right], \\[5pt]
B^c_2 = \left[    \begin{array}{ccc}
         0    &     0   &      0\\
         0    &     0   &      0\\
   45.6448    &     0   &      0\\
         0    &     0   &-1.2500\\
    1.0431    &     0   &      0\\
         0    &     0   &      0\\
         0    &     0   &      0\\
         0    & 0.0183  &  0.0229\\
         0    &     0   &      0\\
   45.6448    &     0   &      0\\
    1.0431    &     0   &      0
\end{array}   \right], ~~
C^c_2 = \left[    \begin{array}{c}
     0  \\
   0.0242  \\
     0\\
  0.0072\\
    0   \\
    0 \\
  -0.0241 \\
  -0.9722 \\
  1.0000 \\
  0.0001 \\
    0
\end{array}   \right]^\mathsf{T}, ~
D^c_2 =  \left[ 0  ~ 0.0193 ~  0.0242 \right].
\end{array}
\end{eqnarray}
}
Open-loop matrices
{\tiny
\begin{eqnarray}
\begin{array}{l}
A^{ol}_2 = 1.0e+004 \\
\left[    \begin{array}{cccccccccccccccc}
    0.0002  & 0.0001 &   0 &   0  &  0.0000   & 0 &  0 &   0  &  0  &  0 &  0 & 0  &  0 & 0 &  0\\
   -0.0001  &  0   &   0  &  0  & -0.0000  &  0  & 0  &  0 & 0  &  0  &  0 &  0  & 0 & 0  &  0  &   0\\
      0   & 0 &  0.0002  &  0.0001  &  0.0000 &  0  & 0 & 0 &  0 &  0 & 0  & 0 & 0 &  0 &   0 &   0\\
      0  &  0 & -0.0001 &  0  & -0.0000 &  0  &   0 &  0 &   0  &  0 &   0 &  0 &  0 &   0  &  0  &  0\\
      0  &  0 & 0  & 0  &  0.0001 & 0 & 0.2404 &  0 & 0.0721 &  0 & 0  &-0.2398&  -9.6753 & 9.9517 & 0.0012 & 0\\
      0  &  0 &   0 &   0  &  0  &  0.0001 &  0  &  0.0000  &  0  &  0  &   0  &  0  &   0  &   0  &  0  &   0\\
      0  &  0 &  0  &   0  &  0  &   0  &  0.0001  &  0  &  0.0000  &  0 &  0  & 0 &  0 &  0  &  0  &  0\\
      0  &  0 &  0  & 0  & 0  & -0.0046  & 0  &  0.0001 &   0  & -0.0001  &    0 &   0   & 0 &  0 &  0  &  0\\
      0  &  0 &  0  & 0  & 0  &0.0001 & -0.0001 & 0.0000 & -0.0000  & 0 &  -0.0000 &  0 &  0 & 0 & 0 &  0\\
      0  & 0  &  0  &  0 &  0  & -0.0001  &  0 &  -0.0000 &  0  & -0.0000 &  0  &  0  &   0 &  0  &  0 &  0\\
      0  &  0 &  0  &   0&  0  &  0  & 0 & 0  &  0  &  0  & -0.0000  &  0  &  0  &  0 &  0 &  0\\
      0  &  0 &   0  &  0&  0  & 0 &   0 &  0  &  0 &  0  &  0 &  -0.0000 &  0  &  0 & 0  &  0\\
      0  & 0  & 0    &0  &  0  &  0 &  0.0000 &  0  & 0.0000  & 0 &  0 &  -0.0000 &  -0.0000 &  0  & 0 &  0\\
      0  & 0  &  0   & 0 &  0  & 0 &  0  &  0 & 0  & 0  &  0  &  0 &  0 &   0.0001 &  0.0000 &   0\\
      0  &  0 &   0  &  0&  0  & 0  & 0  &   0 & 0  &  0 &   0  & 0  & 0 &  -0.0046 &  0.0001 & -0.0001\\
      0  &  0 &  0   & 0 & 0   & 0  &  0 &   0  &   0  & 0  &  0  &  0 &  0 &  -0.0001 & -0.0000 & -0.0000
\end{array}   \right], \\[5pt]
B^{ol}_2 = 1.0e+003 \left[    \begin{array}{cccc}
         0  &       0  &       0  &       0\\
         0  &       0  &       0  &       0\\
         0  &       0  &       0  &       0\\
         0  &       0  &       0  &       0\\
         0  &  1.9228  &  2.4035  & -0.2441\\
         0  &       0  &       0  &       0\\
         0  &       0  &       0  &       0\\
    0.0456  &       0  &       0  &       0\\
         0  &       0  & -0.0013  &       0\\
    0.0010  &       0  &       0  &       0\\
         0  &       0  &       0  &       0\\
         0  &       0  &       0  &       0\\
         0  &  0.0000  &  0.0000  &       0\\
         0  &       0  &       0  &       0\\
    0.0456  &       0  &       0  &       0\\
    0.0010  &       0  &       0  &       0
\end{array}   \right], ~~
C^{ol}_2 =C^{ol}_1,  ~~ D^{ol}_2 =D^{ol}_1, ~~
||G^{ol}_2(z)||_{\infty} < \gamma_2, ~ \gamma_2 =3.0784e+005.
\end{array}
\end{eqnarray}
}
Closed-loop matrix
{\tiny
\begin{eqnarray}
\begin{array}{l}
A^{cl}_2 = 1.0e+004\\
 \left[    \begin{array}{cccccccccccccccc}
  0.0002 & 0.0001  &  0  &   0  &  0.0000 &    0   &  0 &   0   & 0  &  0  &  0   & 0 &  0 &  0  &  0  & 0\\
 -0.0001 &  0  &   0 &    0 &  -0.0000 &   0 &     0  & 0&   0 &   0  &   0  &  0&   0 &   0 &   0 &   0\\
     0   & 0 &  0.0002 &   0.0001 &   0.0000 &   0  &  0  &  0  &  0  &  0   &  0 &    0  &  0  &  0  & 0 &  0\\
     0   & 0 &  -0.0001  & 0  & -0.0000 &   0 &  0   & 0 &  0 &   0  &  0  & 0  & 0  &  0  &  0  &  0\\
     0   & 0 &  -0.1923 & 0  & 0.0001 & 0 &  0.2404 & 0 & 0.0721 & 0 & 0 & -0.2398 & -9.6753 & 9.9517 & 0.0012 & 0\\
      0   &0 &    0   &  0  &  0  &  0.0001 &   0  &  0.0000 &  0  &  0 &   0  &  0  &  0  &  0  & 0  &  0\\
      0   & 0 &   0  &  0  &  0  &  0  &  0.0001 &   0 &  0.0000  &  0  &  0 &   0 &   0  &  0 &  0  &   0\\
 -0.0046  & 0 &   0 &   0  &  0  & -0.0046 &   0  &  0.0001 &  0 &  -0.0001  &  0 &   0 & 0  &   0 &  0 &   0\\
      0   & 0 &    0 & 0  & 0  & 0.0001 &  -0.0001 &   0.0000 & -0.0000 &  0  & -0.0000 & 0 & 0 & 0 & 0 &  0\\
 -0.0001  &  0&    0 &  0 &   0 & -0.0001 & 0  & -0.0000 &  0 &  -0.0000 &  0  &  0  & 0  &  0 &  0 &   0\\
      0   & 0  &  0  &  0  &  0 & 0  &  0  & 0 &  0  &   0  & -0.0000 &   0  &  0  &  0  &  0 &   0\\
      0   & 0  &   0  &   0 &    0 & 0  &  0  &  0 &  0  &  0 &  0 &  -0.0000 &   0  &  0 &   0  &   0\\
      0   & 0  & -0.0000 &   0  & 0 &  0 & 0.0000 &  0 & 0.0000 &  0 & 0  & -0.0000  & -0.0000 &  0 & 0 &  0\\
      0   & 0  &   0  & 0  &  0   &  0  & 0   & 0  &  0  &   0  &   0  &  0  &   0  &  0.0001  &  0.0000  &  0\\
 -0.0046  &  0 &   0  & 0  &  0  & 0  &  0  & 0    & 0 & 0 &  0  & 0 & 0 & -0.0046 &  0.0001  & -0.0001\\
 -0.0001  &  0 &   0  &  0 &   0 &  0  &  0 &   0 &   0 &   0  & 0 &   0 &    0 &  -0.0001 &  -0.0000 &  -0.0000
\end{array}   \right].
\end{array}
\end{eqnarray}
}

\textbf{3- MCM (Maximum Continuous Climb)}\\
Equilibrium values:
$NH_{eq}=49554~(RPM)$, $NL_{eq}=40354~(RPM)$, $Wf_{eq}=95.2~(kg/hr)$, $PLA_{eq}=30~(deg)$. The nondimensional values of the main plant states in the code are $NH_{mcm}=\frac{NH_{eq}}{NH_{ref}}=97.7030 \%$, and $NL_{mcm}=\frac{NH_{eq}}{NH_{ref}}=92.4787 \%$. \\	

Plant matrices
{\footnotesize
\begin{eqnarray}
\begin{array}{l}
A^p_3 = \left[    \begin{array}{cccc}
    1.9746  & 1.0000  &     0    &     0\\
   -0.9747  &    0    &     0    &     0\\
         0  &    0    & 1.9742   & 1.0000\\
         0  &    0    & -0.9742  &      0
\end{array}   \right], ~~
B^p_3 = \left[    \begin{array}{c}
      0.0052\\
     -0.0051\\
      0.0061\\
     -0.0060
\end{array}   \right], ~
C^p_3 = C^p_1, ~
D^p_3 = D^p_1.
\end{array}
\end{eqnarray}
}
Controller matrices
{\scriptsize
\begin{eqnarray}
\begin{array}{l}
A^c_3   = \left[    \begin{array}{cccccccccccc}
    1.0000   &  0  &  0.0001  & 0  &  0  &  0  &  0  & 0  & 0  & 0  &  0\\
     0  &  1.0000 &  0  & 0.3000 &  0  &   0  &  0  & 0  & 0  &  0  &  0\\
  -44.2236 &  0 &  0.7598  &  0  & -0.7606 &  0 &   0  & 0  & 0 & 0 &  0\\
   1.2500  & -1.2500  & 0.0002  & -0.3750  & 0 & -0.0040  & 0 &  0 &  0  & 0 &  0\\
   -1.0375  &  0 &  -0.0056  &  0  & -0.1490  &  0  &  0  &   0  & 0 &  0 &   0\\
      0  & 0  &    0  &   0  &   0 & -0.0566 & 0  &  0  &  0 &  0  &  0\\
      0  & 0  &   0  &   0  &   0  &  0 &  -0.0476  &   0 &  0 &  0  &   0\\
      0  & 0.0235 &  0 &  0.0070 & 0 &  0 &  -0.0234  & -0.1667 &  0  &  0 &  0\\
      0  &  0  &   0  & 0   &  0 & 0  &  0 & 0 & 1.0000 &  0.0001 &  0\\
      0  &  0  &   0  & 0  &   0 &   0 &  0  &  0 & -44.2236  & 0.7598  & -0.7606\\
      0  &  0  &  0  & 0  & 0   &  0 & 0  &   0 & -1.0375 &  -0.0056  & -0.1490\\
\end{array}   \right], \\[5pt]
B^c_3 = \left[    \begin{array}{ccc}
         0  &       0  &       0\\
         0  &       0  &       0\\
   44.2236  &       0  &       0\\
         0  &       0  & -1.2500\\
    1.0375  &       0  &       0\\
         0  &       0  &       0\\
         0  &       0  &       0\\
         0  &  0.0188  &  0.0235\\
         0  &       0  &       0\\
   44.2236  &       0  &       0\\
    1.0375  &       0  &       0\\
\end{array}   \right], ~~
C^c_3 = \left[    \begin{array}{c}
   0  \\
   0.0249 \\
   0 \\
   0.0075 \\
   0  \\
   0 \\
   -0.0248 \\
   -0.9722\\
   1.0000  \\
   0.0001\\
   0
\end{array}   \right]^\mathsf{T}, ~
D^c_3 =  \left[ 0  ~ 0.0199  ~ 0.0249 \right].
\end{array}
\end{eqnarray}
}
Open-loop matrices
{\tiny
\begin{eqnarray}
\begin{array}{l}
A^{ol}_3 = 1.0e+004\\
 \left[    \begin{array}{cccccccccccccccc}
   0.0002 & 0.0001 &   0  &   0 &  0.0000 & 0  &   0 &   0  & 0  & 0  & 0  &  0  &  0  &  0 &  0  &  0\\
  -0.0001 & 0  &  0 &   0 &  -0.0000 &  0  & 0  &  0 & 0 &   0 &   0  & 0  &  0 & 0  &  0 &   0\\
       0  & 0   & 0.0002   & 0.0001  &  0.0000  &  0  &  0  &  0 & 0  &  0  &  0 &  0  &  0 &   0 &   0 &  0\\
       0  & 0  & -0.0001  &  0 &  -0.0000  & 0  & 0  & 0  & 0  &  0  &  0 &  0 &  0  &  0 &   0 &   0\\
       0  & 0  & 0 &  0  &  0.0001 & 0 & 0.2474 &  0 & 0.0742 & 0  & 0 & -0.2468 & -9.6753 & 9.9517 & 0.0013 & 0\\
       0  & 0   & 0 &  0  &  0 & 0.0001  &  0  & 0.0000 & 0  & 0  &  0  & 0  &  0  &   0  & 0  &   0\\
       0  & 0   & 0 &  0  &  0 &  0 &  0.0001 &   0 & 0.0000 &  0  &  0  &   0 &  0 &   0 &  0  &  0\\
       0  & 0   & 0 &  0  &  0 & -0.0044  & 0 &  0.0001 &  0  & -0.0001 &   0 &  0 &   0 &  0 &  0 &  0\\
       0  & 0   & 0 &  0  &  0 &  0.0001  &-0.0001  & 0.0000 & -0.0000 & 0 &  -0.0000 &   0 &   0  &  0 &  0 &  0\\
       0  & 0   & 0 &   0 &  0 & -0.0001   & 0  & -0.0000 & 0 &  -0.0000  &  0  & 0  &  0  &  0  &  0  &   0\\
       0  & 0   & 0 &   0 &  0 &  0  &  0   & 0   &0   & 0  & -0.0000  &  0  & 0 &  0  &  0  &  0\\
       0  & 0   &0  &  0  &  0 &  0  &  0   & 0   &0   & 0   & 0 &  -0.0000  &  0  &   0 &  0  &  0\\
       0  & 0   &0  &  0  &  0 &  0  &  0.0000 &   0  &0.0000 &  0  & 0  & -0.0000 &  -0.0000 &   0 &  0  &  0\\
       0  & 0   & 0 &   0 &  0 &  0  &  0   & 0  &0   & 0  &   0  &  0 &   0 & 0.0001 & 0.0000  &   0\\
       0  & 0   & 0 &   0 &  0 &  0  &  0  & 0  &0   & 0  &  0  &  0  &  0 &  -0.0044  & 0.0001 & -0.0001\\
       0  & 0   & 0 &   0 &  0 &  0  &  0  & 0 & 0  &  0  &  0  &  0 &   0  & -0.0001 &  -0.0000 &  -0.0000
\end{array}   \right], \\[5pt]
B^{ol}_3 = 1.0e+003 \left[    \begin{array}{cccc}
         0  &       0  &       0  &       0\\
         0  &       0  &       0  &       0\\
         0  &       0  &       0  &       0\\
         0  &       0  &       0  &       0\\
         0  &  1.9789  &  2.4736  & -0.2441\\
         0  &       0  &       0  &       0\\
         0  &       0  &       0  &       0\\
    0.0442  &       0  &       0  &       0\\
         0  &       0  & -0.0013  &       0\\
    0.0010  &       0  &       0  &       0\\
         0  &       0  &       0  &       0\\
         0  &       0  &       0  &       0\\
         0  &  0.0000  &  0.0000  &       0\\
         0  &       0  &       0  &       0\\
    0.0442  &       0  &       0  &       0\\
    0.0010  &       0  &       0  &       0
\end{array}   \right], ~~
C^{ol}_3 =C^{ol}_1,  ~~ D^{ol}_3 =D^{ol}_1, ~~
||G^{ol}_3(z)||_{\infty} < \gamma_3, ~ \gamma_3 =3.4046e+004
\end{array}
\end{eqnarray}
}
Closed-loop matrix
{\tiny
\begin{eqnarray}
\begin{array}{l}
A^{cl}_3 = 1.0e+004\\
 \left[    \begin{array}{cccccccccccccccc}
    0.0002 & 0.0001&   0 &   0 & 0.0000  &  0  &  0  &  0 & 0  &  0  &  0  &  0  &  0 &  0  & 0  &  0\\
   -0.0001 &  0    & 0  & 0 & -0.0000  & 0  &  0  & 0 & 0 &  0 &  0 &   0 & 0  & 0 &  0  &  0\\
       0   & 0   & 0.0002&  0.0001   & 0.0000  &  0 &  0  &  0 & 0  &  0  &  0 &  0 &  0 &   0  &  0  &  0\\
       0   &  0  & -0.0001&    0  & -0.0000  &  0 &   0 & 0 & 0  &   0  &  0  &  0 &  0  &  0 &  0  &  0\\
       0  & 0  & -0.1979  &  0  & 0.0001  & 0  & 0.2474 & 0 & 0.0742 & 0 &  0 & -0.2468 & -9.6753 & 9.9517 & 0.0013 & 0\\
       0   & 0  &   0  & 0  & 0  & 0.0001   & 0 &  0.0000 &  0  &  0  &   0 &   0  & 0  & 0  &  0 &   0\\
       0   & 0  &   0  & 0  &  0  & 0   & 0.0001  &  0 & 0.0000 &  0 &   0 &   0 &0  & 0  &  0 &  0\\
  -0.0044  &  0 &   0  &  0 &   0 & -0.0044 &    0  &  0.0001 & 0 & -0.0001  &  0 &   0  &  0  &   0 &  0  &  0\\
       0   & 0  &  0   & 0  &  0  & 0.0001 &  -0.0001 & 0.0000 & -0.0000  &  0  & -0.0000 &   0 &0  &  0 &0  &  0\\
   -0.0001 &  0 &  0   & 0  &  0  &-0.0001 &   0 & -0.0000 & 0 & -0.0000  &  0 &   0 &  0  &  0  &  0  &   0\\
       0   & 0  &  0   &  0 &   0 &   0 &  0  &  0 &  0  &  0  & -0.0000  &  0 &    0  &  0  &  0  &   0\\
       0   & 0  &  0   &  0 &   0 &   0  &   0 &    0  & 0  &  0  &  0  & -0.0000  & 0  &  0 &   0 &   0\\
       0   & 0  & -0.0000 &  0  &  0  &  0  &  0.0000 &  0 & 0.0000  &  0 &  0 &  -0.0000 &  -0.0000 &   0 &0  & 0\\
       0   &  0 &   0 &   0  &  0 &  0 &  0  &  0  & 0  &  0   & 0  & 0   & 0 &   0.0001 &  0.0000  &   0\\
  -0.0044  &  0 &   0 &   0  &  0 &  0 &  0  &  0 & 0   &  0   &  0  &   0 &  0 &  -0.0044 &  0.0001  & -0.0001\\
  -0.0001  &  0 &   0 &   0  & 0  &  0  & 0  &  0 &  0  &  0  &  0   & 0 &  0 &  -0.0001 &  -0.0000 &  -0.0000
\end{array}   \right].
\end{array}
\end{eqnarray}
}

\textbf{4- TOP (Take-Off Power)}\\
Equilibrium values:
$NH_{eq}=51230~(RPM)$, $NL_{eq}=43902~(RPM)$, $Wf_{eq}=112.4~(kg/hr)$, $PLA_{eq}=40~(deg)$. The nondimensional values of the main plant states in the code are $NH_{top}=\frac{NH_{eq}}{NH_{ref}}=101.0075 \%$, and $NL_{top}=\frac{NH_{eq}}{NH_{ref}}=100.6096 \%$. \\

Plant matrices
{\footnotesize
\begin{eqnarray}
\begin{array}{l}
A^p_4 = \left[    \begin{array}{cccc}
    1.9746  & 1.0000  &     0    &     0\\
   -0.9747  &    0    &     0    &     0\\
         0  &    0    & 1.9742   & 1.0000\\
         0  &    0    & -0.9742  &      0
\end{array}   \right], ~~
B^p_4 = \left[    \begin{array}{c}
      0.0045\\
     -0.0045\\
      0.0056\\
     -0.0056
\end{array}   \right], ~
C^p_4 = C^p_1, ~ D^p_4 = D^p_1.
\end{array}
\end{eqnarray}
}
Controller matrices
{\scriptsize
\begin{eqnarray}
\begin{array}{l}
A^c_4   = \left[    \begin{array}{cccccccccccc}
    1.0000 &   0  & 0.0001  &   0   &  0  &  0  &   0  & 0  &  0  &   0 &    0\\
     0   & 1.0000 &   0  & 0.3000  & 0  &  0 &   0  &  0  & 0  & 0  &  0\\
  -41.8743 &  0  & 0.7565 &  0  & -0.6932  &  0  & 0  & 0 &  0  &   0  &  0\\
   1.2500 &  -1.2500 &   0.0002 &  -0.3750 &  0  & -0.0040 &   0 &  0 &   0 & 0 &  0\\
  -1.0265 &  0  & -0.0060  &  0  & -0.1457  &  0  &  0  &  0 &  0  &   0 &  0\\
     0   & 0    & 0   &  0  &  0 &  -0.0566  &  0  &  0 &  0  &  0 &   0\\
     0   &  0   & 0   & 0   & 0  &  0  & -0.0476 &  0 &  0 &    0  &   0\\
     0   & 0.0245 &   0 &   0.0074  &   0  &    0 &  -0.0245  & -0.1667 &  0  &  0  &  0\\
     0   &  0     &0    & 0  & 0  &  0  &   0   &   0  & 1.0000  &  0.0001   &   0\\
     0   &  0     &0    & 0  &  0   &  0  &   0    &  0  &-41.8743 &   0.7565 &  -0.6932\\
     0   &  0     &0    & 0  &   0  &   0  &   0  &   0  &-1.0265 &  -0.0060  & -0.1457
\end{array}   \right], \\[5pt]
B^c_3 = \left[    \begin{array}{ccc}
         0  &       0 &        0\\
         0  &       0 &        0\\
   41.8743  &       0 &        0\\
         0  &       0 &  -1.2500\\
    1.0265  &       0 &        0\\
         0  &       0 &        0\\
         0  &       0 &        0\\
         0  &  0.0196 &   0.0245\\
         0  &       0 &        0\\
   41.8743  &       0 &        0\\
    1.0265  &       0 &        0
\end{array}   \right], ~~
C^c_4 = \left[    \begin{array}{c}
    0 \\
    0.0261 \\
      0 \\
    0.0078 \\
      0 \\
      0 \\
    -0.0260 \\
  -0.9722\\
  1.0000 \\
   0.0001 \\
    0
\end{array}   \right]^\mathsf{T}, ~
D^c_4 =  \left[ 0 ~  0.0209 ~ 0.0261 \right].
\end{array}
\end{eqnarray}
}

Open-loop matrices
{\tiny
\begin{eqnarray}
\begin{array}{l}
A^{ol}_4 = 1.0e+004 \\[5pt]
\left[    \begin{array}{cccccccccccccccc}
  0.0002& 0.0001 & 0 &  0 &  0.0000  &  0  &  0  & 0  & 0  &  0  & 0 &  0  &  0  &  0 &  0 &  0\\
 -0.0001&  0  &  0  &  0  & -0.0000  &  0  &  0  &  0 &  0 &  0  & 0  & 0  &  0  &  0  & 0  & 0\\
     0  &  0  &  0.0002 & 0.0001 &  0.0000 &  0 &   0 &  0 &  0  & 0  & 0  &  0  &  0  & 0  & 0 &   0\\
     0  &  0  & -0.0001 & 0  & -0.0000  & 0  & 0 &  0 &  0 &  0  & 0  &  0 &   0 &   0 &  0 &  0\\
     0  &  0  &  0  &  0 &  0.0001 & 0 &  0.2598 &  0 & 0.0779  & 0 &  0  & -0.2592&  -9.6753  & 9.9517 & 0.0014  & 0\\
     0  &  0  &  0  &  0 &  0 &  0.0001 &   0 &  0.0000 &  0    & 0  &   0  &  0 &   0 &   0 &  0 &  0\\
     0  &  0  &  0  &  0 &  0 &  0  & 0.0001 &  0 & 0.0000 &  0  &  0 &   0 &  0  &  0  &  0 &  0\\
     0  &  0  &  0  &  0 &  0 &  -0.0042 &   0  & 0.0001  & 0  & -0.0001  &  0 &  0  &  0 &   0 &  0  &  0\\
     0  &  0  &  0  &  0 &  0 &  0.0001 & -0.0001&  0.0000 & -0.0000 &   0 &  -0.0000&  0 &    0 &   0 &  0 &   0\\
     0  &  0  &  0  &  0 &  0 &  -0.0001  &  0  & -0.0000 & 0 &  -0.0000  & 0 &   0 &  0 &   0 &   0  &  0\\
     0  &  0  &  0  &  0 &  0 &   0  &  0  &  0  & 0  &  0 &  -0.0000  &  0  &  0  &  0  &  0  &  0\\
     0  &  0  &  0  &  0 &  0 &   0  &  0  &  0 &  0 &  0  &  0 &  -0.0000 & 0  &  0 &  0  &  0\\
     0  &  0  &  0  &  0 &  0 &   0  &  0.0000 &  0  &0.0000  &  0  &  0 & -0.0000  & -0.0000 &  0  &  0 &  0\\
     0  &  0  &  0  &  0 &  0 &   0  &  0 &  0 &  0  &  0  &  0  &  0  &  0 &  0.0001  & 0.0000 &  0\\
     0  &  0  &  0  &  0 &  0 &   0  &  0 &  0 & 0   &  0  & 0   & 0  &   0 &  -0.0042 &   0.0001 &  -0.0001\\
     0  &  0  &  0  &  0 &  0 &   0  &  0 & 0  & 0   & 0   & 0   & 0  &  0  & -0.0001 &  -0.0000  & -0.0000
\end{array}   \right], \\[5pt]
B^{ol}_4 = 1.0e+003 \left[    \begin{array}{cccc}
         0  &       0 &        0  &       0\\
         0  &       0 &        0  &       0\\
         0  &       0 &        0  &       0\\
         0  &       0 &        0  &       0\\
         0  &  2.0782 &   2.5977  & -0.2441\\
         0  &       0 &        0  &       0\\
         0  &       0 &        0  &       0\\
    0.0419  &       0 &        0  &       0\\
         0  &       0 &  -0.0013  &       0\\
    0.0010  &       0 &        0  &       0\\
         0  &       0 &        0  &       0\\
         0  &       0 &        0  &       0\\
         0  &  0.0000 &   0.0000  &       0\\
         0  &       0 &        0  &       0\\
    0.0419  &       0 &        0  &       0\\
    0.0010  &       0 &        0  &       0
\end{array}   \right], ~~
C^{ol}_4 =C^{ol}_1,  ~~ D^{ol}_4 =D^{ol}_1, ~~
||G^{ol}_4(z)||_{\infty} < \gamma_4, ~ \gamma_4 =9.2748e+004
\end{array}
\end{eqnarray}
}

Closed-loop matrix
{\tiny
\begin{eqnarray}
\begin{array}{l}
A^{cl}_4 = 1.0e+004 \\[5pt]
\left[    \begin{array}{cccccccccccccccc}
    0.0002 & 0.0001  &  0  &  0  & 0.0000  &  0  &  0  & 0  & 0 &   0  &  0  & 0  &  0  &   0 &   0 &   0\\
   -0.0001 & 0 &  0  &  0  & -0.0000 &   0 &   0  &   0 &  0  &   0  &  0  &   0 &    0 &  0  &  0 &    0\\
     0   & 0 & 0.0002 &  0.0001 &  0.0000 &   0  &  0  &  0  & 0  &  0  &0  &0  &  0 &   0 &   0 &  0\\
     0   & 0 &  -0.0001 &    0  & -0.0000  &  0  &  0  &   0  & 0  &  0 &  0  & 0  & 0  &0   &  0 &   0\\
     0   & 0 &  -0.2078 &  0 &  0.0001 & 0  & 0.2598 &  0  & 0.0779 &  0 &  0 & -0.2592 & -9.6753 & 9.9517& 0.0014 & 0\\
     0   & 0 &  0   &  0 &  0  &  0.0001 &   0  & 0.0000 & 0  &   0  &  0  &0 &   0  &  0  &0  &  0\\
     0   & 0 &   0  &   0  &  0 &   0 &  0.0001 &  0 & 0.0000 &   0  &  0 &   0 &    0 &   0 &   0  &  0\\
  -0.0042&  0&   0  &  0  &  0 &  -0.0042  & 0 &  0.0001  & 0  & -0.0001 &  0  &   0 &   0 &   0  &0  & 0\\
     0   & 0 &   0  &  0  &  0 &  0.0001 & -0.0001 &  0.0000 & -0.0000 &  0 &  -0.0000 &  0  &  0  & 0  &  0 &  0\\
   -0.0001&   0 &   0&    0 &   0  & -0.0001  & 0 & -0.0000  & 0 &  -0.0000 &  0  &  0  &  0  &  0 &  0  &   0\\
     0   & 0   &  0  &  0  & 0    & 0   &  0  &  0 &  0  &  0 &  -0.0000  &  0  &  0   & 0 &  0 &   0\\
     0   & 0   &  0  &  0  & 0   &  0   & 0  &  0  &  0  &  0  &   0  & -0.0000 &   0 &   0  &   0  &  0\\
     0   & 0   &-0.0000 &  0&   0  & 0  & 0.0000 &  0 & 0.0000 &  0  &  0  & -0.0000  & -0.0000 &  0 &  0 &  0\\
     0   & 0   & 0  &  0    & 0 &   0 &   0  & 0  & 0  &  0  &  0  &  0  &  0  &  0.0001  & 0.0000  &  0\\
   -0.0042 &  0 &0   &0  &  0 &   0  &   0   &  0  & 0   &  0  &  0  &  0  &  0 &  -0.0042  & 0.0001 &  -0.0001\\
   -0.0001 &  0 &0   &0  &0 &  0  &  0  & 0 & 0   &  0 &  0  &  0 &  0  & -0.0001 &  -0.0000  & -0.0000
\end{array}   \right].
\end{array}
\end{eqnarray}
}


\end{document}